\documentclass[10pt,aps,pre,amsmath,amsfonts,amssymb,floatfix,longbibliography,superscriptaddress,notitlepage, twocolumn]{revtex4-1}

\usepackage{mathrsfs} \usepackage{graphicx,color} \usepackage{bbm} \usepackage{multirow}
\usepackage[normalem]{ulem}
\usepackage{lineno}
\usepackage{empheq}

  \def\erf{\text{erf}}

\newcommand{\red}[1]{\textcolor{black}{#1}}

\def\to{\rightarrow}

\newcommand{\beq}{\begin{equation}} \newcommand{\eeq}{\end{equation}}

\renewcommand{\phi}{\varphi}

\renewcommand{\vec}[1]{\boldsymbol{\mathrm{#1}}}
\def\vr{{\boldsymbol r}}

\begin{document}

\title{Determining the nonequilibrium criticality of a Gardner transition via a hybrid study of molecular simulations and machine learning}

\author{Huaping Li}
\altaffiliation{Contributed equally to this work}
\affiliation{School of Chemistry, Beihang University, Beijing 100191, China}
\affiliation{Center of Soft Matter Physics and Its Applications, Beihang University, Beijing 100191, China}
\affiliation{Wenzhou Institute, University of Chinese Academy of Sciences, Wenzhou, Zhejiang 325000, China}

\author{Yuliang Jin}
\altaffiliation{Contributed equally to this work}
\email{yuliangjin@mail.itp.ac.cn}
\affiliation{CAS Key Laboratory of Theoretical Physics, Institute of Theoretical Physics, Chinese Academy of Sciences, Beijing 100190, China}
\affiliation{School of Physical Sciences, University of Chinese Academy of Sciences, Beijing 100049, China}

\author{Ying Jiang}
\email{yjiang@buaa.edu.cn}
\affiliation{School of Chemistry, Beihang University, Beijing 100191, China}
\affiliation{Center of Soft Matter Physics and Its Applications, Beihang University, Beijing 100191, China}

\author{Jeff Z. Y. Chen}
\affiliation{Department of Physics and Astronomy, University of Waterloo, Waterloo, Ontario, Canada N2L 3G1}

\begin{abstract}
Apparent critical phenomena, typically indicated by  growing correlation lengths and dynamical slowing-down, 
are  ubiquitous in non-equilibrium systems such as supercooled liquids, amorphous solids, active matter and spin glasses.
It is often challenging to determine if such observations are related to a true  second-order phase transition as in the equilibrium case,  or simply a crossover, and even more so to measure the associated critical exponents. 
\red{Here, we show that the simulation results of a hard-sphere glass in three dimensions, 
are consistent with the recent theoretical prediction of a Gardner transition, a continuous non-equilibrium  phase transition.}
Using a hybrid molecular simulation - machine learning approach, we obtain scaling laws for both finite-size and aging effects, and determine the critical exponents that traditional methods fail to estimate.
Our study provides a novel approach that is useful to understand the nature of glass transitions, and can be generalized to analyze  other non-equilibrium phase transitions.
\end{abstract}

\maketitle
\def\thefootnote{*}\footnotetext{These authors contributed equally to this work}\def\thefootnote{\arabic{footnote}}

Among all transitions in glassy systems, the Gardner transition is perhaps the most peculiar one, considering its remarkably complex way to break the symmetry~\cite{gardner1985spin,  charbonneau2014fractal, berthier2019gardner, charbonneau2017glass}. 
According to the mean-field theory that is exact in large dimensions, it is a second-order phase transition separating the {\it simple glass phase} and the {\it Gardner phase} where the free energy basin splits into many marginally stable sub-basins~\cite{charbonneau2014fractal}. In structural glasses, the Gardner transition 
occurs deep in the glass phase below the  liquid-glass transition temperature, which is observable even under non-equilibrium conditions~\cite{charbonneau2015numerical, berthier2016growing, seoane2018spin, seguin2016experimental, geirhos2018johari, Hammond5714,  jin2018stability, jin2017exploring, liao2019hierarchical},
and has important consequences on the rheological and mechanical properties of the material~\cite{biroli2016breakdown, jin2018stability, jin2017exploring}, as well as 
on the jamming criticality at zero temperature~\cite{charbonneau2015jamming}.
From a theoretical viewpoint,  the Gardner transition universality class contains other important cases such as the famous de Almeida-Thouless transition in spin glasses~\cite{de1978stability}.



As a non-equilibrium, continuous phase transition, the Gardner transition is expected to display the divergence of (i) the fluctuations of the caging order parameter that characterizes the particle vibrations~\cite{charbonneau2015numerical, berthier2016growing}, (ii) the length scale for the spatial correlation between individual cages~\cite{berthier2016growing}, and (iii) the time scale to reach the {\it restricted equilibrium}~\cite{rainone2015following} deep in the glass phase.
Previous  computer simulations of hard-sphere glasses in $d=2$~\cite{liao2019hierarchical}  and $d=3$ dimensions~\cite{berthier2016growing, seoane2018spin}, 
and experiments of molecular glass formers~\cite{geirhos2018johari}, granular~\cite{seguin2016experimental} and colloidal~\cite{Hammond5714} glasses, 
showed consistent evidence for above signature features.
However, whether or not the ``Gardner transition'' is a true phase transition
in physical dimensions remains hotly  debated: it has been argued that
the transition could be eliminated by critical finite-dimensional fluctuations and local defects~\cite{hicks2018gardner,scalliet2017absence, urbani2015gardner}, but a recent field-theory calculation up to the three-loop expansion indeed found fixed points even below the upper critical dimension $d_{\rm u} = 6$~\cite{charbonneau2017nontrivial}. 
To our knowledge, there have been no reliable measurements of the critical exponents of the Gardner transition neither from  simulations nor from experiments. 


In this paper, we \red{aim to examine whether}
the Gardner transition \red{satisfies characteristic scalings of a}  second-order phase transition in a three-dimensional computer simulated hard-sphere glass.
We propose a scaling ansatz for the {\it caging susceptibility}~\cite{charbonneau2015numerical, berthier2016growing} in the Gardner phase, which combines the logarithmic aging behavior~\cite{seoane2018spin} and the standard critical finite-size scaling. We further determine the values of two independent critical exponents, which are in line with previous theoretical predictions~\cite{charbonneau2017nontrivial}.
In particular, the exponent $\nu$ for the correlation length
is obtained by a machine learning approach~\cite{Carrasquilla2017Machine,Nieuwenburg_NatPhys2017}, which is shown to be able to capture  the hidden features of simple glass/Gardner phases from the massive data set generated by molecular simulations.

\begin{figure*}[ht]
\centerline{\includegraphics[width=0.85\textwidth]{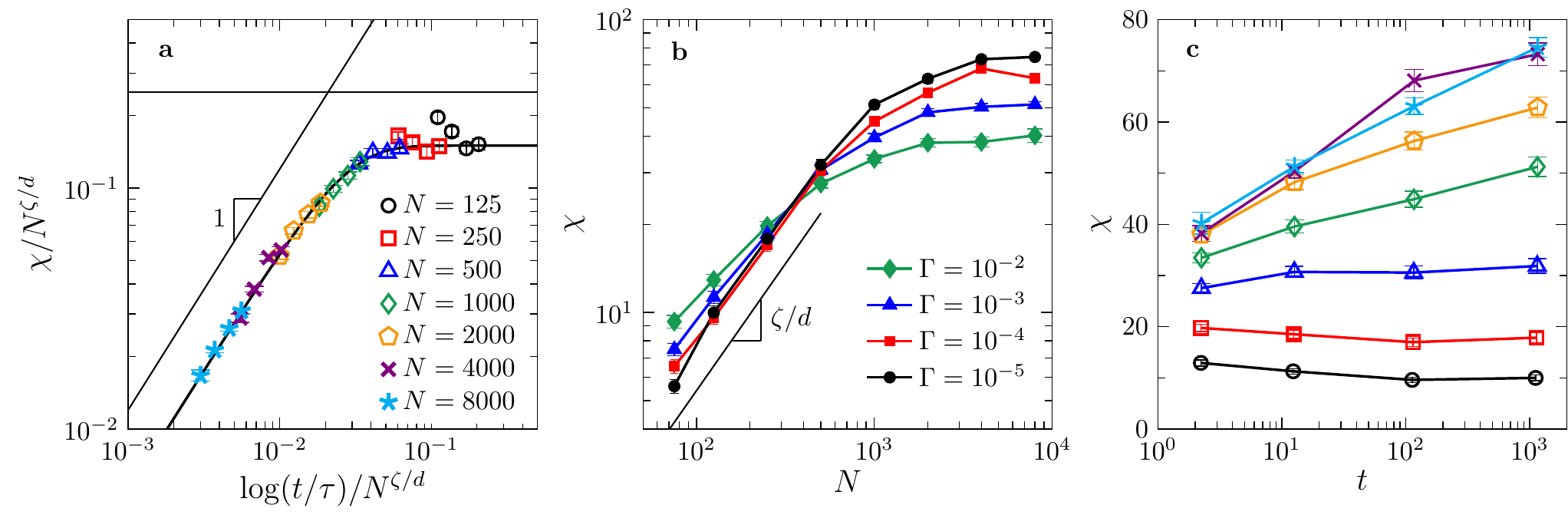}}
\caption{{\bf
Finite-size and aging scalings of caging susceptibility in the Gardner phase.}
Data are obtained for a fixed $\hat{T}=0.00385$ below $\hat{T}_{\rm G}$ using $N_{\rm r} = 5$ glass replicas for each equilibrium sample, and are averaged over $N_{\rm s} = 480$ equilibrium samples.
(a)
Susceptibility data collapsed according to the scaling function Eq.(~\ref{eq:ansatz2}), for $125 \leq N \leq 8000$ and $\Gamma = 10^{-2}, 10^{-3}, 10^{-4},10^{-5}$, where  the parameters $\zeta = 2.6$ and $\tau = 0.0016$ are determined independently as shown in Fig.~\ref{fig:T-dependence}.
The lines represent asymptotic behaviors $\mathcal{F}(x\to \infty) \sim 1$ and $\mathcal{F}(x\to 0) \sim x$.
We also show 
an empirical fitting using the hyperbolic tangent  function, $\mathcal{F}(x) = 0.15 \tanh(37x)$.
(b) Susceptibility as a function of system size $N$, for a few different quench rate $\Gamma$. The line indicates the finite-size scaling $\chi \sim N^{\zeta/d}$ (see Eq.~\ref{eq:chi_L}).
(c) Susceptibility as a function of quench time $t$, for a few different $N$. The same data are plotted in (a-c), and the legend in (a) applies to both (a) and (c). Error bars represent the standard error of the mean in all figures. }
\label{fig:scaling}
\end{figure*}


\subsection*{Results}

We simulate a polydisperse hard-sphere glass model in $d=3$ dimensions (see {\it Materials and Methods}). An efficient Monte-Carlo swap algorithm \red{\cite{grigera2001fast, berthier2016equilibrium}}  (see {\it Materials and Methods}) is employed to prepare dense equilibrium samples at a (reduced) temperature $\hat{T}_{\rm g} = 0.033$ (or volume fraction $\varphi_{\rm g} = 0.63$; $\hat{T}$ and $\varphi$ are related through equations of states, see Fig.~S1 in {\it Supporting Information Appendix}), which is below the mode-coupling theory (MCT) temperature
$\hat{T}_{\rm MCT} \approx 0.044$ (or $\varphi_{\rm MCT} \approx 0.594$)~\cite{berthier2016growing}.
Glass configurations are generated by  quenching (compressing) the system from $\hat{T}_{\rm g} $ to various target $\hat{T}$, with  a constant quench (compression)  rate $\Gamma$, using the Lubachevsky-Stillinger algorithm (see {\it Materials and Methods}).
The quench (compression) time $t \propto 1/\Gamma$  plays a similar role as the waiting time (or aging time) after rapid quenching~\cite{seoane2018spin}.
Previous simulations suggest that the system undergoes  a Gardner crossover around  $\hat{T}_{\rm G} \approx 0.0078$ (or $\varphi_{\rm G} \approx 0.67$) for the given  $\hat{T}_{\rm g} = 0.033$, in systems of $N=1000$ particles~\cite{berthier2016growing}. Jamming occurs at  the zero temperature limit $\hat{T} \to 0$ (or $\varphi_{\rm J} \approx 0.682$) ~\cite{berthier2016growing}, where particles form an isostatic contact network.

\begin{figure}[ht]
\centerline{\includegraphics[width=0.5\textwidth]{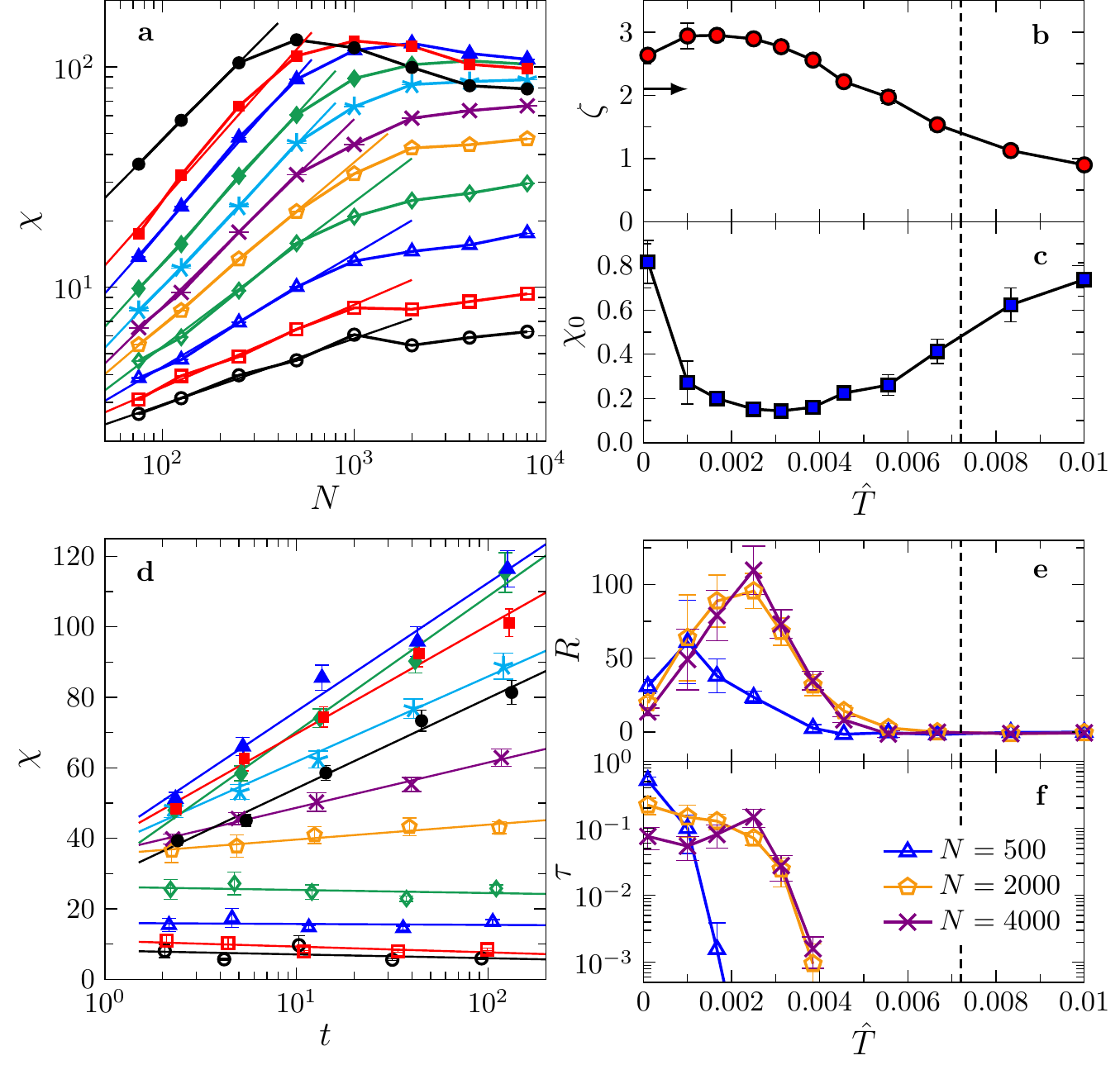}}
\caption{
{\bf Temperature dependence of finite-size and aging scalings of caging susceptibility.}
(a) Susceptibility $\chi$ as a function of $N$, for a fixed $\Gamma=10^{-4}$ and  $\hat{T} = 1/100$, 1/120, 1/150, 1/180, 1/220, 1/260, 1/320, 1/400, 1/600, 1/1000, 1/10000 (from bottom to top). Data are obtained by using $N_{\rm r}=5$ glass replicas and are
averaged over $N_{\rm s} = 1200$ equilibrium samples. The data points in the power-law regime are fitted to Eq.~(\ref{eq:chi_L}) (lines), and the fitting parameters $\zeta(\hat{T})$ and $\chi_0(\hat{T})$ are plotted in (b) and (c). The  theoretical exponent $\zeta = 2.1$~\cite{charbonneau2017nontrivial} is marked by the horizontal arrow in (b).
(d) Susceptibility $\chi$ as a function of $t$, for a fixed $N=4000$ and a few different  $\hat{T}$ ($N_{\rm r}=5$, $N_{\rm s}=240$, see panel (a) and its caption for the values of $\hat{T}$). The data are fitted to Eq.~(\ref{eq:chi_t}) (lines), and the fitting parameters $R(\hat{T})$ 
and $\tau(\hat{T})$ are plotted in (e) and (f), where we have used the values of $\chi_0(\hat{T})$ plotted in (c).
The data for $N=500$ and $N=2000$ are also plotted in (e-f) to show that the behavior of the curves becomes $N$-independent in sufficiently large systems, within the numerical errors.
The Gardner transition temperature  $\hat{T}_{\rm G} = 0.0072$ (see Fig.~\ref{fig:ML}) is indicated by the vertical dashed lines in (b-c, e-f).}
\label{fig:T-dependence}
\end{figure}

\begin{figure}[ht]
\centerline{\includegraphics[width=0.5\textwidth]{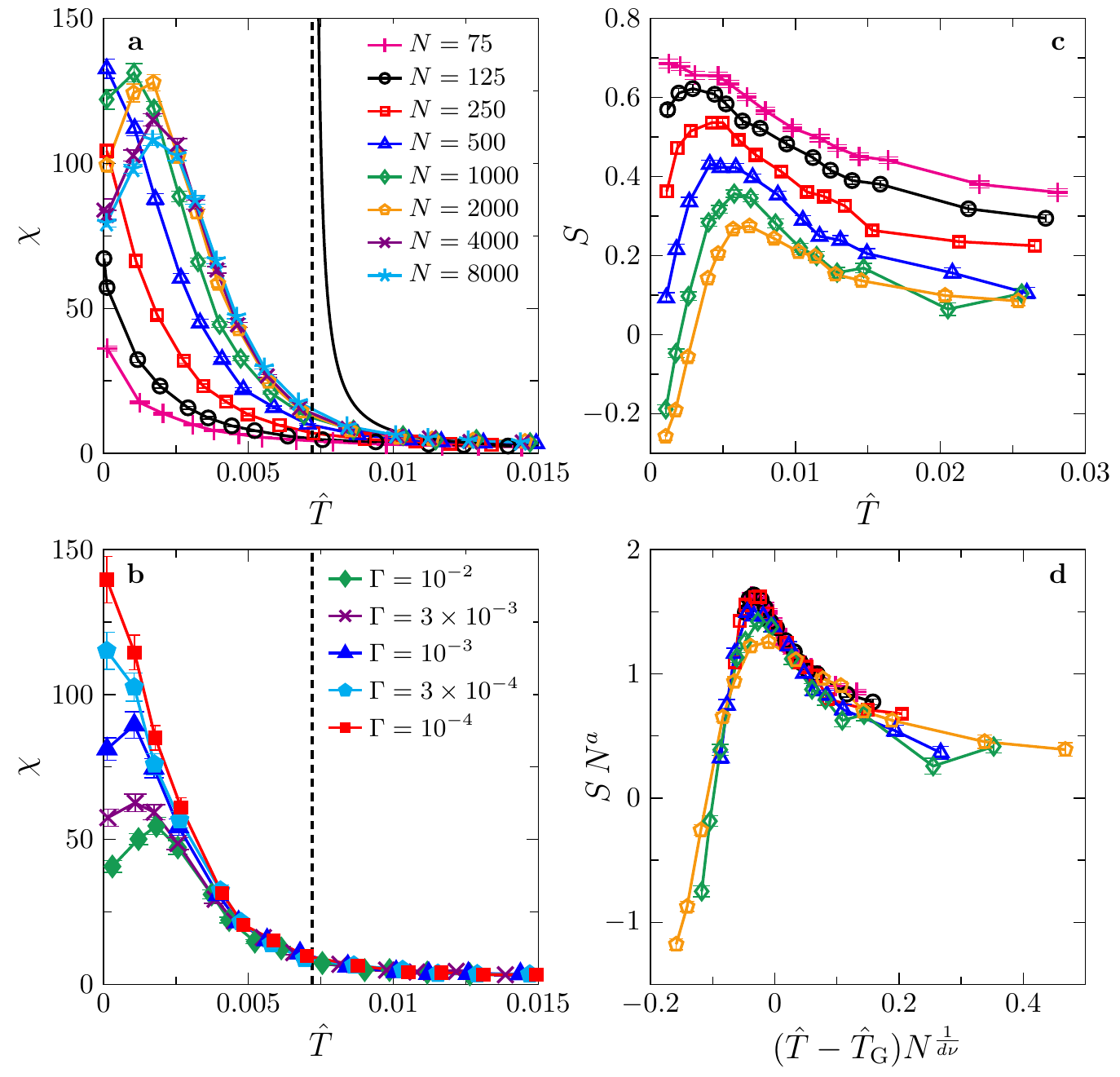}}
\caption{{\bf Examining the criticality of the Gardner transition via the data of caging susceptibility and skewness.}
Susceptibility $\chi$ is plotted as a function of $\hat{T}$ ($N_{\rm r} = 5$ and $N_{\rm s} = 1200$), for (a) a fixed $\Gamma = 10^{-4}$ and a few different $N$, and (b) a fixed $N=500$ and a few different $\Gamma$.
To demonstrate how far the data are away from the critical scaling, we plot a line in (a) representing $\chi \sim (\hat{T} - \hat{T}_{\rm G})^{-\gamma}$, where we set $\gamma  = \zeta \nu \approx 1.2$, estimated from $\zeta \approx 1.5$ (see Fig.~\ref{fig:T-dependence}b) and $\nu = 0.78$ (see Fig.~\ref{fig:ML}h).
(c) Skewness $S$ as a function of $\hat{T}$ for a few different $N$ ($N_{\rm r} = 10$ and $N_{\rm s} = 2400$). (d) Data collapsing according to the scaling ansatz $S N^a \sim \mathcal{S}\left[ (\hat{T} - \hat{T}_{\rm G})  N^{\frac{1}{d \nu}}\right]$, where $a=0.2$ is a fitting parameter, and the values $\hat{T}_{\rm G} = 0.0072$ and $\nu = 0.78$ are obtained from the machine learning method (see Fig.~\ref{fig:ML}). The vertical dashed lines in (a-b) mark  $\hat{T}_{\rm G}$.
The legend in (a) applies to (a, c-d).
 }
\label{fig:criticality}
\end{figure}

The  static correlation length of  the Gardner transition is predicted to diverge at the transition point from above~\cite{charbonneau2014fractal},
\red{
 \beq
\xi_{\rm s} (\hat{T}) \sim 
\begin{cases}
(\hat{T} - \hat{T}_{\rm G} )^{-\nu}, & {\rm for}\,\, \hat{T} > \hat{T}_{\rm G};\\
\infty, & {\rm for}\,\, \hat{T} \leq \hat{T}_{\rm G}.
\end{cases}
\label{eq:xi_infty}
\eeq
}
Different from a standard second-order phase transition, here $\xi_{\rm s}$ diverges not only at, but also below the transition point, since the system in the entire Gardner phase is marginally stable.
Moreover, such a static correlation length is only reached 
\red{in restricted equilibrium when the aging effects disappear.}
Note that we only consider aging attributed to the Gardner transition, not to the glass transition (or $\alpha$-processes) \red{~\cite{rainone2015following, berthier2016growing}}.
The $\alpha$-relaxation time  $\tau_\alpha \sim 10^{10}$ at $\hat{T}_{\rm g} = 0.033$ \red{~\cite{berthier2017configurational}}, which would further increase with decreasing $\hat{T}$, is clearly beyond our simulation time window $t \lesssim10^3$~\cite{rainone2015following, berthier2016growing}.
Near or below $\hat{T}_{\rm G}$,
the correlation length is time-dependent at short times due to the aging effects. 
\red{Based on numerical observations, we  propose that the correlation length $\xi(\hat{T}, t)$  follows the following form,
\beq
\xi(\hat{T}, t) =
\begin{cases}
\xi_{\rm d}(\hat{T}, t) = \{ R(\hat{T}) \log [t/ \tau(\hat{T} )] \}^{1/\zeta_{\rm d}}, \,\,  {\rm for} \,\, t < \tau_{\rm G}(\hat{T});\\
\xi_{\rm s}, \,\,  {\rm for} \,\,  \tau_{\rm G}(\hat{T})< t < \tau_{\alpha},
\end{cases}
\label{eq:xi}
\eeq
where $R(\hat{T})$, $\tau(\hat{T} )$ and $\zeta_{\rm d}$ are parameters to be determined.
The static correlation length  $\xi_{\rm s}$ is defined in Eq.~(\ref{eq:xi_infty}), to be distinguished from 
 the dynamical correlation length  $\xi_{\rm d}(\hat{T}, t)$.
Here $\tau_{\rm G}(\hat{T})$ is the time scale associated with the Gardner transition, which becomes large near and below $\hat{T}_{\rm G}$~\cite{berthier2016growing}. Note that $\tau_{\rm G}$ is always smaller than $\tau_\alpha$, for  the system to remain in the glass state.}
The logarithmic aging behavior \red{in Eq.~(\ref{eq:xi})} has been observed in many non-equilibrium systems, including
rapidly quenched hard-sphere glasses~\cite{seoane2018spin} and  spin glasses~\cite{bouchaud1998out}, and is
consistent with the droplet theoretical picture~\cite{fisher1988nonequilibrium}.
\red{
Critical aging (a power-law growth of susceptibility or correlation length) 
 is not observed in our simulation data (see Fig.~\ref{fig:scaling}), in agreement  with an earlier study~\cite{seoane2018spin}.  Ref.~\cite{seoane2018spin} also reports, similarly,  the absence of power-law aging in a three-dimensional spin glass under an external field.} 

\begin{figure*}[ht]
\centerline{\includegraphics[width=\textwidth]{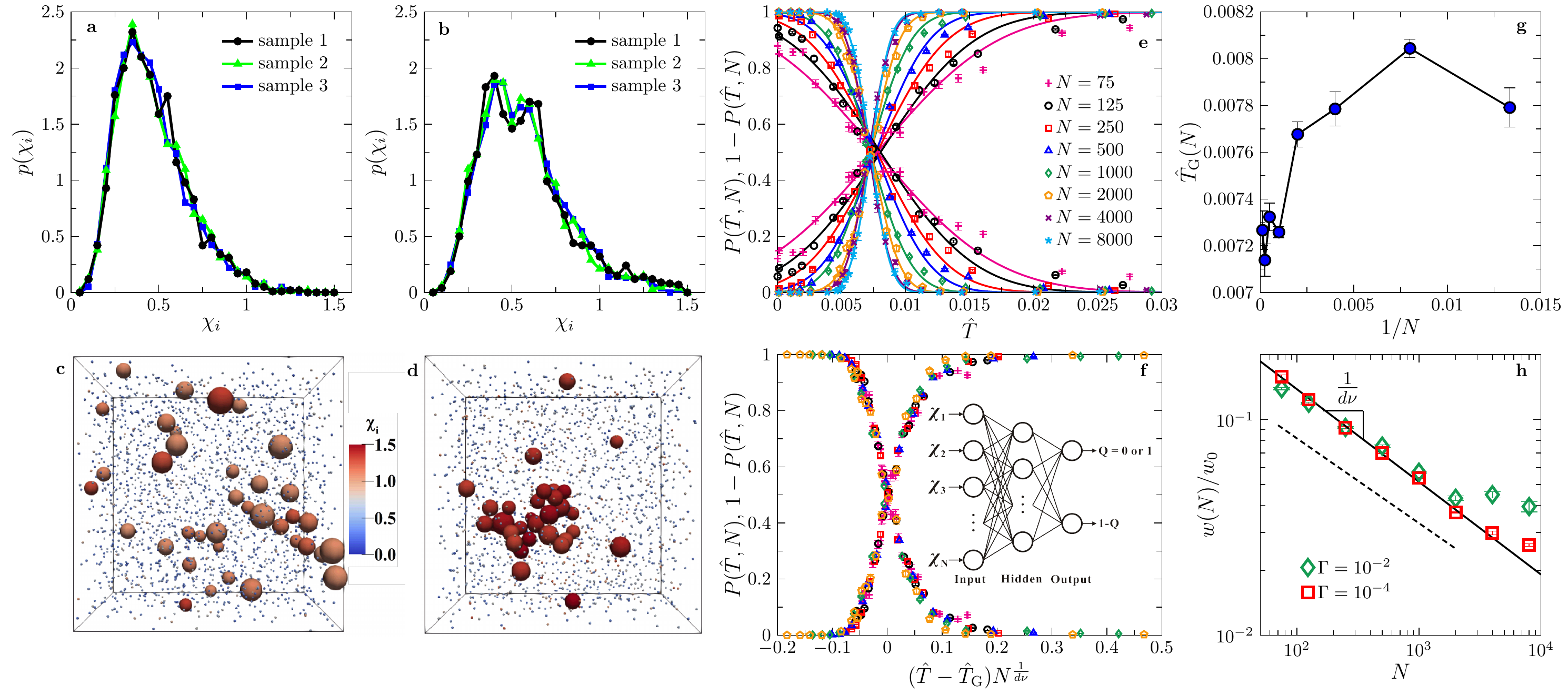}}
\caption{{\bf Machine learning the Gardner transition}.
(a) Probability distribution $p(\chi_i)$ at $\hat{T} = 0.025$ (above $\hat{T}_{\rm G}$) and $\hat{T} = 0.001$ (below $\hat{T}_{\rm G}$) for three different samples, with $N=2000$.  
Spatial distributions of the $\sim 2\%$ particles with the largest $\chi_i$ in sample 1 (other particles are represented by points) are visualized  in (c) for $\hat{T} = 0.025$ and (d) for $\hat{T} = 0.001$,  
which show the difference on caging heterogeneity.
(c) Probabilities $P(\hat{T}, N)$ and $1- P(\hat{T}, N)$ obtained from the machine learning method  are  plotted as functions of (e) $\hat{T}$ and (f) $[\hat{T}- \hat{T}_{\rm G}(N)] N^{\frac{1}{d \nu}}$, for a few different $N$ and $\Gamma = 10^{-4}$. The lines in (e) represent fitting to an  empirical form
$P(\hat{T}, N) = \frac{1}{2} + \frac{1}{2}\erf \left\{\left[\hat{T} - \hat{T}_{\rm G}(N)\right]/w(N) \right\}$, where $\erf(x)$ is the error function. The fitting parameters $\hat{T}_{\rm G}(N)$ and $w(N)$ are plotted in (g) and (h).
The asymptotic transition temperature $\hat{T}_{\rm G} \equiv \hat{T}_{\rm G} (N \to \infty)= 0.0072(2)$ is estimated from (g).
The line in (e) represents
fitting according to the critical scaling $w(N) = w_0 N ^{-\frac{1}{d \nu}}$ within the range $N\leq 2000$, which gives $\nu = 0.78(2)$. 
The shifted data for $\Gamma = 10^{-2}$ are also plotted, which show a narrower critical scaling regime. The theoretical exponent $\nu = 0.85$~\cite{charbonneau2017nontrivial}  is indicated by the dashed line.
The inset of (f) shows a schematic of the FNN architecture.}
\label{fig:ML}
\end{figure*}

While the direct estimate of  the correlation length is technically difficult~\cite{berthier2016growing}, the above scalings are useful in understanding 
the behavior of other important quantities, such as the 
caging susceptibility $\chi$, which  characterizes the fluctuation of the caging order parameter and can be \red{measured} in simulations (see {\it Materials and Methods}).
\red{The divergence of susceptibility is one of the  characteristics of a continuous phase  transition.  
Near and below $\hat{T}_{\rm G}$, 
because $\tau_{\rm G}$ is extremely large,
it becomes impratctical to directly obtain samples in restricted equilibrium.
Thus, one needs to generalize the standard finite-size scaling analysis for equilibrium systems, into a combined finite-size-finite-time scaling analysis, in order to derive critical parameters from 
out of (restricted) equilibrium
data. Such an approach has been developed in Ref.~\cite{lulli2016out} for spin glasses, except that
here a logarithmic (instead of a power-law) growth form of correlation length is used  (see Eq.~\ref{eq:xi}).}

\red{According to the renormalization group theory, close to the critical point, the caging susceptibility should obey the following scaling function~\cite{hohenberg1977theory, lulli2016out},
\beq
\frac{\chi(\hat{T}, L, t)}{\chi_0(\hat{T}) L^{\zeta_{\rm s}}} = \mathcal{F}\left \{ \left [ \frac{\xi(\hat{T}, t)}{L} \right ]^{\zeta_{\rm s}} \right \},
\label{eq:ansatz}
\eeq
where $\chi_0(\hat{T})$ is a temperature-dependent parameter, $L=N^{1/d}$ is the linear size of the system, and $\zeta_{\rm s}$ is the exponent for the static finite-size scaling.
Equation~(\ref{eq:ansatz}) is a strong assertion  that a single, universal scaling can connect the behavior of  caging susceptibility  in the aging regime  (see Eq.~\ref{eq:xi}) to that in the restricted equilibrium regime (see Eq.~\ref{eq:xi_infty}).  The former is dominated by activated dynamics as considered in the droplet theory, while the latter is described by the Gardner transition physics.
The general function form, $\mathcal{F}(x)$, beyond the two dynamical regimes discussed below, was not determined previously.}

\red{(I) In the restricted equilibrium regime ($\tau_{\rm G} < t < \tau_{\alpha}$), $\xi(\hat{T}, t)$ converges to the static correlation length $\xi_{\rm s}$.  In order to recover,  from Eqs.~(\ref{eq:xi_infty}) and~(\ref{eq:ansatz}), the standard scaling of susceptibility in large systems ($\xi_{\rm s}/L \ll 1$),
\beq
\chi(\hat{T}, L) \sim (\hat{T} - \hat{T}_{\rm G})^{-\gamma},
\label{eq:chi_static}
\eeq
where $\gamma = \nu \zeta_{\rm s}$, we require that asymptotically $\mathcal{F}(x \to 0) \sim x$.}

\red{(II) In the aging regime ($t < \tau_{\rm G}$), only the dynamical correlation length $\xi_{\rm d}$ is relevant. Two scalings can be further derived.}

\red{(IIa) For small systems with  $L \ll \xi(\hat{T}, t)$, $\chi$ should be determined by $L$ and  independent of $t$, following the standard finite-size scaling,
\beq
\chi(\hat{T}, L, t) \sim \chi_0(\hat{T}) L^{\zeta_{\rm s}},
\label{eq:chi_L}
\eeq
which requires that  $\mathcal{F}(x \to \infty) \sim 1$.}

\red{
(IIb) For large systems with $L \gg \xi(\hat{T}, t)$, since $\mathcal{F}(x \to 0) \sim x$, Eq.~(\ref{eq:ansatz}) gives, 
\beq
\chi(\hat{T}, L, t)  \sim \chi_0(\hat{T}) \left\{ R(\hat{T}) \log [t/ \tau(\hat{T}) ] \right\}^\kappa.
\label{eq:chi_t}
\eeq
where $\kappa = \zeta_{\rm s}/\zeta_{\rm d}$. In general, the dynamical exponent $\zeta_{\rm d}$ and the static exponent $\zeta_{\rm s}$ do not have to be identical. In spin glasses, $\kappa$ was found to be close to one ($\kappa \in[1,2]$)
~\cite{jonsson2002domain}. In this study, we use the simplest assumption, $\kappa=1$, to capture our simulation results (see Fig.~\ref{fig:scaling}).
Under this assumption, 
we will use a single exponent $\zeta =  \zeta_{\rm s} = \zeta_{\rm d}$ in following analyses.
}


To examine above  \red{expected scalings}, 
 we first consider the case for a fixed $\hat{T}$ below  $\hat{T}_{\rm G}$ where aging clearly presents. Under this condition,
using Eq.~(\ref{eq:xi}) we can simplify
Eq.~(\ref{eq:ansatz}) into the form (the $\hat{T}$-dependence is omitted since $\hat{T}$ is fixed), 
\beq
\frac{\chi (L, t)}{L^\zeta} \sim \mathcal{F}\left[ \frac{\log(t/\tau)}{L^\zeta} \right],
\label{eq:ansatz2}
\eeq
which is confirmed by the numerical data in Fig.~\ref{fig:scaling}(a).
The finite-size scaling Eq.~(\ref{eq:chi_L}) is supported by the data in Fig.~\ref{fig:scaling}(b) for small $N$, while breakdowns are observed for larger $N$ implying the violation of the condition $L \ll \xi(\hat{T}, t)$.
 The scaling regime expands with decreasing $\Gamma$ (or increasing $t$), as the correlation length grows with time.
At even larger $N$, the susceptibility  approaches to a constant value, suggesting that the other asymptotic limit $L \gg \xi(\hat{T}, t)$ has been reached and therefore the value of susceptibility  is determined by $\xi(\hat{T}, t)$ instead of $L$.
The logarithmic growth Eq.~(\ref{eq:chi_t}) is consistent with the data in
Fig.~\ref{fig:scaling}(c) for large $N$,
while in small systems, the susceptibility is independent of $t$, implying $L \ll \xi(\hat{T}, t)$. The scalings are robust with respect to protocol parameters (see Fig.~S3) and the aging protocol (see Fig.~S4).


We next investigate how the parameters in scalings Eqs.~(\ref{eq:chi_L}) and~(\ref{eq:chi_t}) depend on $\hat{T}$.
Fitting data at different $\hat{T}$, obtained from a slow quench rate $\Gamma = 10^{-4}$,  to Eq.~(\ref{eq:chi_L}) in the scaling regime (Fig.~\ref{fig:T-dependence}a), 
gives the value of exponent $\zeta$, which depends weakly on $\hat{T}$ (Fig.~\ref{fig:T-dependence}b).
For $\hat{T} \leq \hat{T}_{\rm G}$,  $\zeta$ is in a range $\sim[1.5, 3.0]$, which is comparable with the theoretical prediction $\zeta = 2.1$~\cite{charbonneau2017nontrivial} (see Table~S1). 
In order to obtain a more accurate estimate of  $\zeta$, one must further decrease $\Gamma$ so that the scaling regime can be extended (see Fig.~\ref{fig:scaling}b), which is unfortunately beyond the present computational power (recall that aging is logarithmically slow). 
The pre-factor $\chi_0(\hat{T})$ behaves non-monotonically with $\hat{T}$, showing a growth  approaching  $\hat{T} = 0$,
which suggests  a stronger finite-size effect in the jamming limit (Fig.~\ref{fig:T-dependence}c).
The value of $\chi_0(\hat{T})$ is in the same order of the individual caging susceptibility (Fig.~S2a), consistent with the interpretation of $\chi_0(\hat{T})$ as the small-$L$ limit of $\chi$, according to Eq.~(\ref{eq:chi_L}).
Figure~\ref{fig:T-dependence}a also shows that the power-law regime shrinks as $\hat{T} \to 0 $.
Because the finite-size scaling only holds when $L \ll \xi(\hat{T}, t)$, it implies that $\xi(\hat{T}, t)$, with $t$ fixed, decreases near the jamming limit, which is confirmed by the direct measurement of  $R(\hat{T})$ (see Fig.~\ref{fig:T-dependence}e and related discussions).
At low $\hat{T}$ and large $N$ (e.g., $\hat{T} = 10^{-4}$ and $N > 500$), the susceptibility slightly decreases with $N$, instead of staying as a constant. 
This effect might be due to a higher-order correction term $L^{-\omega}$ to the scaling function Eq.~(\ref{eq:ansatz}), as has been observed similarly in spin glasses~\cite{banos2012thermodynamic}, but we do not further discuss it here.


Figure~\ref{fig:T-dependence}d shows how the aging scaling Eq.~(\ref{eq:chi_t}) depends on $\hat{T}$.
The aging effect is negligible, i.e., $R(\hat{T})\sim 0$,  above $T_{\rm G}$ (Fig.~\ref{fig:T-dependence}e), consistent with previous  observations based on dynamics of the caging order parameter~\cite{berthier2016growing}.
The non-monotonic behavior of  $R(\hat{T})$ in Fig.~\ref{fig:T-dependence}e can be understood from the mixed impacts from two transitions: aging emerges as $\hat{T}$ lowered below the Gardner transition $\hat{T}_{\rm G}$, which however should naturally  slow down when approaching the jamming transition limit $\hat{T} \to 0 $ where all dynamics freeze.
Accordingly, the susceptibility $\chi$ should also change non-monotonically with $\hat{T}$ in sufficiently  large systems (Fig.~\ref{fig:criticality}a). Interestingly, a very similar non-monotonic behavior of $\chi$ has been reported for the three-dimensional  Edwards-Anderson spin-glass model in an external magnetic field~\cite{seoane2018spin}. 


So far we have discussed the behavior of the  susceptibility and correlation length in the aging regime (Eq.~\ref{eq:xi}). 
In the following we analyze the  restricted equilibrium regime, 
aiming to  examine the criticality near the Gardner transition by estimating the transition temperature $\hat{T}_{\rm G}$ and in particular the exponent $\nu$ in Eq.~(\ref{eq:xi_infty}). 
However, conventional approaches fail to achieve the goal, for the following reasons.
(i) Due to the limited system sizes that can be obtained in simulations, extracting the correlation length  from fitting the correlation function is difficult~\cite{seoane2018spin}. (ii) 
\red{The scaling Eq.~(\ref{eq:chi_static})}
is unobservable in our data (Fig.~\ref{fig:criticality}a-b), suggesting that the systems are too small and the condition $L \gg \xi_{\rm s}$ for the scaling is not satisfied in the critical regime. (iii) In standard second-order phase transitions, the Binder parameter $B(\hat{T}, L)$ (see {\it Materials and Methods})
is independent of the system size  
at the critical temperature. However, $B(\hat{T}, L)$ for different $L$ measured in our simulations do not cross at $\hat{T}_{\rm G}$ (see Fig.~S6), due to the asymmetry of the order parameter distribution as indicated by the non-zero value of the skewness $S$ (see {\it Materials and Methods} for the definition and Fig.~\ref{fig:criticality}c for the data). 
The same reason prevented locating the de Almeida-Thouless transition by the Binder parameter in spin glasses, previously~\cite{ciria1993ahneida}.

To overcome the difficulties, we develop a machine learning approach (see {\it Materials and Methods} and Sec.~S3) using a  feedforward neural network (FNN), inspired by a recent work~\cite{Carrasquilla2017Machine}. 
The method was shown to be able to correctly capture the criticality of phase transitions in several equilibrium systems, including the standard $d=2$ Ising model~\cite{Carrasquilla2017Machine}. Here we generalize it to non-equilibrium phase transitions.
Because the Gardner transition is not accompanied by any obvious structural ordering~\cite{charbonneau2017glass}, a naive 
attempt to train the neural network  based on static configurations fails to learn the transition.
 Instead, we utilize the replica method~\cite{mezard1987spin, parisi2020theory, berthier2016growing} to construct  {\it single-particle caging susceptibilities} $\{ \chi_i \}$ (see {\it Materials and Methods} and Sec.~S3A) as the input data, 
 which encode the change of particle vibrational features 
 around the Gardner transition. Indeed, the distribution probability $p(\chi_i)$  displays a distinction above and below $\hat{T}_{\rm G}$, showing single- and double-peaks respectively (Figs.~\ref{fig:ML}a-b), which is accompanied consistently by the difference on vibrational heterogeneity~\cite{berthier2016growing} (Figs.~\ref{fig:ML}c-d).

 
 Once well trained, the FNN output layer provides a
  probability $P(\hat T, N)$ of an $N-$particle system belonging to the Gardner phase at $\hat T$ (correspondingly $1-P$  represents the probability in the simple glass phase, see Fig.~\ref{fig:ML}e). 
The finite-size analysis according to the scaling invariance  $P(\xi_{\rm s}/L) \sim P[(\hat{T} - \hat{T}_{\rm G})N^{\frac{1}{d\nu}}]$ (see Eq.~\ref{eq:xi_infty}) can give both the transition temperature $\hat T_G$ and critical exponent $\nu$. 
This strategy  is standard in the analysis of continuous phase transitions such as a percolation transition -- the difference is that 
it is straightforward to identify a percolated configuration without the need to use machine learning.
\red{In Sec.~S6, we show that the machine learning method can be used to pin-down the critical temperature $T_{\rm c}$  and the correlation length exponent $\nu$ of the spin glass transition in a $d=3$ spin glass model.}


The asymptotic critical temperature is estimated to be 
$\hat T_{\rm G} \equiv\hat T_{\rm G}(N\rightarrow\infty)=0.0072(2)$ from the data obtained by $\Gamma = 10^{-4}$ (Fig.~\ref{fig:ML}g), or equivalently $\varphi_{\rm G}=0.670(1)$, which is consistent with the previous independent measurement~\cite{berthier2016growing}. 
Fitting the width $\omega(N)$ of $P(\hat T,N)$  to the scaling $\omega(N) \sim N^{-\frac{1}{d\nu}}$, in the range $N\leq N^*\approx 2000$, gives $\nu=0.78(2)$ (Fig.~\ref{fig:ML}h), which is close to the  theoretical prediction $\nu=0.85$~\cite{charbonneau2017nontrivial} (see Table~S1). Here $N^*$ is the cutoff size beyond which the critical scaling does not hold.
Consequently, using the estimated $\hat{T}_{\rm G}$ and $\nu$,
the data of $P[(\hat{T} - \hat{T}_{\rm G})N^{\frac{1}{d\nu}}]$ for  different $N$ with $N \leq N^*$ collapse onto a universal master curve (Fig.~\ref{fig:ML}f).
The machine learning results are further confirmed by the collapse of skewness data using the scaling $S(\hat{T}, N ) N^a \sim \mathcal{S}\left[ (\hat{T} - \hat{T}_{\rm G})  N^{\frac{1}{d \nu}}\right]$, with a fitted exponent $a=0.2$ (Fig.~\ref{fig:criticality}d and Fig.~S5).

To better understand the meaning of $N^*$, it is useful to re-examine the susceptibility data near $\hat{T}_{\rm G}$.
For a fixed $\Gamma = 10^{-4}$, the finite-size effect disappears when $N>N^*\approx 2000$ (Fig.~\ref{fig:criticality}a), suggesting that the aging effect (Eq.~\ref{eq:xi}) becomes dominant. 
 On the other hand, for a fixed $N=500 < N^*$, $\chi$ is independent of $\Gamma$ below $10^{-2}$, implying that further decreasing $\Gamma$  would not change the scaling in such small systems. 
 Therefore, only systems with $N \leq N^*$ would follow the correct finite-size critical scaling. Very importantly, the cutoff size $N^*$, and thus the critical scaling regime, extends (Fig.~\ref{fig:ML}h, Fig.~\ref{fig:criticality}a and Figs.~S12-13) with decreasing $\Gamma$, \red{which indicates a growing correlation length as $\Gamma \to 0$}. 

\subsection*{Discussion}
\red{The finite-size and finite-time analyses performed in this study, facilitated by a machine learning method, show the critical behavior of a Gardner transition in a hard-sphere glass model. It should be pointed out that the size of the simulated system, as well as the observed critical scaling regime,  is limited. We thus}
cannot exclude the possibility that the correlation length is finite but larger than the maximum $L$ simulated in this study. \red{
Considering that about 8 million core-hours were used for this work, 
studying larger systems is, unfortunately,  beyond the current computational power.}
 
 \red{As a non-equilibrium phase transition, the discussion  of the Gardner physics shall be restricted within the life-time $\tau_\alpha$ of the glass sample.
Because in finite dimensions, $\tau_\alpha$ would only diverge 
at the conjectured ideal glass transition temperature $\hat{T}_{\rm K}$,  in principle the Gardner transition can be a true phase transition with a diverging time scale only at $\hat{T}_{\rm G} (\hat{T}_{\rm g} = \hat{T}_{\rm K})$ (note that $\hat{T}_{\rm G}$ is a function of $\hat{T}_{\rm g}$, see Refs.~\cite{charbonneau2014fractal, berthier2016growing}), in the glasses quenched from a glass transition temperature at $\hat{T}_{\rm g} = \hat{T}_{\rm K}$. }

There is a long debate on the nature of the spin glass phase in finite dimensions~\cite{RuizLorenzo2020}. It remains unclear if 
a de Almeida-Thouless transition presents in finite-dimensional spin glasses in a field~\cite{RuizLorenzo2020, baity2014three, banos2012thermodynamic}: strong finite-size effects were observed in the analysis of correlation length, because the measurements are dominated by atypical samples~\cite{baity2014three, parisi2012numerical}. Machine learning approaches~\cite{munoz2020learning} could provide new ideas and opportunities to tackle the problem. For example, since an explicit measurement of correlation length is not required anymore, will the contributions of rare samples be suppressed in the data analysis?
Finally, we point out the possibility to generalize  the method presented here to study phase transitions in other non-equilibrium, disordered systems, including polymer dissolutions~\cite{miller2003review} and 
cells~\cite{hyman2014liquid}. \\ \\

\centerline{\bf METHODS}


\subsection*{Glass model}

The polydisperse hard-sphere model used here has been extensively studied recently~\cite{berthier2016equilibrium, berthier2016growing, jin2018stability, jin2017exploring, seoane2018spin}.
The system consists of $N$ hard spheres in a periodic simulation box of volume $V$, where the particle diameters are distributed according to a continuous function $P_D (D_{\rm min} \leq D \leq D_{\rm min}/0.45) \sim D^{-3}$.
The system is characterized by volume fraction $\varphi$ and reduced temperature $\hat{T} =  1/\hat{P} = Nk_{\rm B}T/PV $, where $P$ is the pressure, $\hat{P}$ the reduced pressure, $k_{\rm B}$  the Boltzmann constant (set  to unity), and $T$  the temperature (set to unity).
In this study, all results are reported in terms of the reduced temperature $\hat{T}$, and  ``reduced" is omitted in the rest of discussions for simplicity. 
The mean diameter $D_{\rm mean}$ and the particle  mass $m$  are used as the units  of length and mass. 
We do not observe any crystallization  during our simulations due to the large polydispersity.

We denote by $\hat{T}_{\rm g}$ the glass transition temperature where the system falls out of equilibrium.
The glass transition temperature $\hat{T}_{\rm g}$ and density  $\varphi_{\rm g}$ are related through  the liquid equation of state (see  Fig.~S1). Glass configurations are created by compressing the system from $\hat{T}_{\rm g}$  to a target  $\hat{T} <\hat{T}_{\rm g}$. The temperature $\hat{T}$ and density $\varphi$ of glasses are related by the glass equation of state (see Fig.~S1)~\cite{berthier2016growing}.
While  in previous studies, the volume fraction  $\varphi$ was more commonly used as the control parameter~\cite{berthier2016growing}, here we instead choose to control $\hat{T}$  in order to  mimic isothermal aging procedures that are widely conducted in experiments. Because by definition  $\hat{P} = 1/\hat{T}$, the reduced pressure is also a constant during aging.

As shown previously, the Gardner transition temperature  $\hat{T}_{\rm G}$ depends on the glass transition temperature $\hat{T}_{\rm g}$~\cite{charbonneau2014fractal, berthier2016growing}. In this study we focus on $\hat{T}_{\rm g} = 0.033$ (or $\varphi_{\rm g} = 0.63$) as a case study, in order to  minimize the unwanted $\alpha-$relaxation processes~\cite{berthier2016growing}, and in the meanwhile to explore as large as possible the ranges of $N$ and $t$,  within our simulation time window.

For each system size $N$  = 75, 125, 250, 500, 1000, 2000, 4000 and 8000, we prepare $N_{\rm s} \sim 2400$ independent samples of equilibrium states  at $\hat{T}_{\rm g} = 0.033$, using the swap algorithm~\cite{grigera2001fast,berthier2016equilibrium}. 
Compared to previous studies~\cite{berthier2016growing, seoane2018spin, liao2019hierarchical} where $N_{\rm s} \sim 100$, a lot more samples are generated, which is essential for the machine learning study.
Each equilibrium state is then compression quenched to $\hat{T}<\hat{T}_{\rm g}$, using the Lubachevsky-Stillinger algorithm~\cite{lubachevsky1990geometric, skoge2006packing}.
To avoid confusion, we call equilibrium states at $\hat{T}_{\rm g}$ as {\it equilibrium samples}, and the quenched configurations at $\hat{T} < \hat{T}_{\rm g}$ as {\it glass replicas}. 
For each equilibrium sample, $N_{\rm r} = 5 -20$
 glass replicas are generated. 
 The $N_{\rm r}$ glass replicas share the same initial particle positions at $\hat{T}_{\rm g}$ given by the equilibrium sample before quenching, but they are assigned by different initial particle velocities  drawn independently from the 
Maxwell-Boltzmann distribution, which yield different configurations at $\hat{T} < \hat{T}_{\rm g}$ after quenching.\\


\subsection*{Protocol to prepare initial configurations - swap algorithm}

The initial configurations at $\hat{T}_{\rm g}$  are prepared by using a swap algorithm~\cite{grigera2001fast, berthier2016equilibrium}.
At each swap Monte Carlo step, two randomly chosen particles are swapped if they do not overlap with other particles at the new positions.
Such non-local Monte Carlo moves, combined  with event-driven molecular dynamics~\cite{jin2017exploring, jin2018stability} or regular Monte Carlo moves~\cite{berthier2016equilibrium}, significantly facilitate the equilibration procedure. \\

\subsection*{Compression protocol - Lubachevsky-Stillinger algorithm}

To simulate the compression quench procedure, the Lubachevsky-Stillinger algorithm~\cite{lubachevsky1990geometric, skoge2006packing} is employed. The algorithm is based on event-driven molecular dynamics.
Starting from an equilibrium configuration at $\hat{T}_{\rm g}$, the  algorithm mimics compression
by inflating particle sizes  with a fixed rate  $\Gamma = \frac{1}{2D} \frac{dD}{dt}$, where the simulation time is expressed in units of $\sqrt{1/k_{\rm B} m  D_{\rm mean}^2}$. 
{The quench time $t$ is the total time used to compress the system from $\hat{T}_{\rm g}$ (where $t=0$) to the target $\hat{T}$ (after quenching, the system is relaxed for a short period of time $t_{\rm w}=1$).}\\

\subsection*{Caging order parameter and cumulants}

The {\it caging order parameter} $\Delta_{AB}$, which characterizes  the average size of particle vibrational cages,  is defined as the mean-squared distance between two glass replicas $A$ and $B$ of the same equilibrium sample~\cite{charbonneau2015numerical, berthier2016growing, seoane2018spin, liao2019hierarchical, scalliet2017absence},
\beq
\Delta_{AB} = \frac{1}{N} \sum_{i=1}^N \left| \vr_i^A - \vr_i^B \right |^2.
\eeq
The caging susceptibility $\chi$, skewness $S$, and Binder parameter $B$ correspond to the second, third, and fourth cumulants of the reduced order parameter $u = \frac{\Delta_{AB} - \langle  \Delta_{AB}  \rangle }{\langle \Delta_{AB} \rangle}$ (note that $\langle u \rangle = 0$ by definition),
\beq
\chi = N \overline{ \langle u^2 \rangle},
\eeq
\beq
S = \overline{\left(\frac{\langle u^3 \rangle}{\langle u^2 \rangle^{\frac{3}{2}}} \right)},
\eeq
and
\beq
B =1 -  \frac{1}{3} \overline{\left(\frac{\langle u^4 \rangle}{\langle u^2 \rangle^{2}} \right)},
\eeq
where $\langle x\rangle$ represents the average over $N_{\rm r}(N_{\rm r}-1)/2$ pairs of glass replicas,
 and   $\overline{x}$ represents the average over $N_{\rm s}$ different initial equilibrium samples (disorder). The contributions from sample-to-sample fluctuations are not included in these definitions (see Sec.~S2A).

The caging order parameter of a single particle $i$ is  $\Delta_{AB}^ i = \left| \vr_i^A - \vr_i^B \right |^2$, and the corresponding reduced parameter is
$u_{i}  = \frac{\Delta_{AB}^i - \langle \Delta_{AB}^i \rangle}{\langle \Delta_{AB}^i \rangle}$ (by definition  $\langle u_{i} \rangle = 0$). The {\it single-particle caging susceptibility} is defined as
\beq
\chi_i = \langle u_i^2 \rangle - \langle u_i \rangle ^2 = \frac{\langle (\Delta_{AB}^i)^2 \rangle - \langle \Delta_{AB}^i \rangle ^2 }{\langle \Delta_{AB}^i \rangle ^2}.
\eeq
Figure~S2 shows that  the average single-particle caging susceptibility $\chi_{\rm ind}$, compared to the total susceptibility $\chi$, is negligible in the Gardner phase, where
the spatial correlations between single-particle caging order parameters dominate. \\

\subsection*{Machine learning  algorithm}

Supervised learning is performed on a FNN, which is 
 composed of one input layer of $N$ nodes, one hidden layer of 128 nodes with exponential linear unit (ELU) activation functions,
  and one output layer providing binary classifications through softmax activation functions.
We adopt the cross-entropy cost function 
with an additional 
L2 regularization term to avoid overfitting. The Adam algorithm is used to implement a stochastic optimization.  
  
  
 


For each system size $N$, we choose $N_{\rm s}^{\rm train}   =  200 - 2000$ independent equilibrium samples to create the training data set.
Each sample  is characterized by an array of  single-particle caging susceptibilities $\chi_1,\chi_2, \cdots, \chi_{N} $ at a given $\hat{T}< \hat{T}_{\rm g}$, which are calculated from $N_{\rm r} = 5$ glass replicas and fed into the FNN as the input data.

During training, the algorithm learns ``hidden features"  of the two phases, by pre-assuming that,  if $\hat{T} > \hat{T}_1 = 0.011$ (or $\hat{T} < \hat{T}_2 = 0.0045$), the input data belong to the simple glass (or the Gardner) phase. 
The parameters $\hat{T}_1$ and $\hat{T}_2$ are preset such that $\hat{T}_2 < \hat{T}_{\rm G} < \hat{T}_1$, 
with the vicinity of $\hat{T}_{\rm G}$ blanked out
(see Sec.~S3C for more details).
Training data are generated at $N_{\hat{T}}$ different temperatures, where $N_{\hat{T}} = 5-6$ in the simple glass phase ($\hat{T} > \hat{T}_1$) and  $N_{\hat{T}} = 6-7$ in the Gardner phase ($\hat{T} < \hat{T}_2$). To effectively expand the training data set, we further apply $ N_{\rm shuffle} = 20-200$ random shuffles  to the array  $\chi_1,\chi_2, \cdots, \chi_{N} $ (see Sec.~S3D). In total, $N_{\rm s}^{\rm train} \times N_{\hat{T}}  \times N_{\rm shuffle}  \sim   10^5$ input arrays in each phase are fed into the FNN.   In Secs.~S3D-F, we discuss in detail the influence of above parameters on the results.

Once trained, the FNN is used in the phase identification of the test data set that contains $ N_{\rm s}^{\rm test} = 40 -  400$  additional samples.
For each test sample $k$ at a temperature $\hat{T}$, the FNN provides a  binary output $Q_k = 1 $ or 0. The probability $P$ of the system being in the Gardner phase is estimated as $P = \frac{1}{N_{\rm s}^{\rm test}}\sum_{k=1}^{N_{\rm s}^{\rm test}} Q_k$ (note that $1-P$ is the probability of being in the simple glass phase). 

We perform 10 independent runs to obtain both the mean and the statistical error of $P(\hat{T}, N)$ as shown in Fig.~\ref{fig:ML}b. 
For each run, $N_{\rm s}^{\rm train}$ training samples and $N_{\rm s}^{\rm test}$ test samples are randomly chosen from the pool of $N_{\rm s}$ total samples generated by molecular simulations, and there is no overlapping between  the training set and the test set. Additional details related to the machine learning method can be found in Sec.~S3.



\begin{acknowledgments}
We warmly thank Patrick~Charbonneau, Beatriz~Seoane,  Qianshi~Wei,  Xin~Xu, Sho~Yaida, Hajime~Yoshino, Francesco~Zamponi~and Haijun~Zhou for inspiring discussions.
We 
 acknowledge funding from
Project 11935002,
Project 11974361, Project 11947302, Project  21622401 \red{and Project 22073004} supported by NSFC, from Key Research Program of Frontier  Sciences, CAS, Grant NO. ZDBS-LY-7017, from 111 Project (B14009), and from NSERC. H. Li is grateful for funding support from the China Postdoctoral Science Foundation (2018M641141).
 This work was granted access to
  the HPC Cluster of ITP-CAS.
 \end{acknowledgments}
 






\clearpage

\centerline{\bf Supplementary Information}

\setcounter{figure}{0}  
\setcounter{equation}{0}  
\setcounter{table}{0} 
\renewcommand\thefigure{S\arabic{figure}}
\renewcommand\theequation{S\arabic{equation}}
\renewcommand\thesection{S\arabic{section}}
\renewcommand\thetable{S\arabic{table}}

\section{Liquid and glass equations of state}
\begin{figure}[ht]
\centerline{\includegraphics[width=0.35\textwidth]{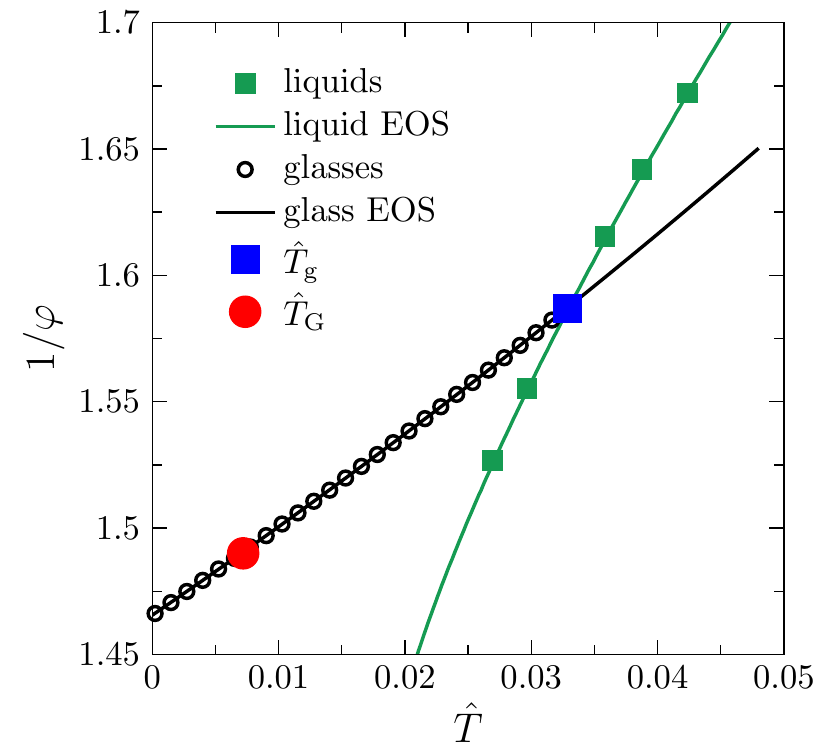}}
\caption{Liquid and glass ($\hat{T}_{\rm g} = 0.033$) EOSs (data adapted from Ref.~\cite{berthier2016growing}). The simulation data are fitted to the empirical Carnahan-Starling liquid EOS~\cite{berthier2016growing} (green line), and  the glass EOS Eq.~(\ref{eq:glass_EOS}) with fitting parameters $c= 1.59$ and $\varphi_{\rm J} = 0.682$ (black line).}
\label{fig:EOS}
\end{figure}

The reduced temperature $\hat{T}$ and the volume fraction $\varphi$ of equilibrium states are related by the liquid equation of state (EOS), as shown in Fig.~\ref{fig:EOS}. The glass EOS depends on the glass transition temperature $\hat{T}_{\rm g}$ that is protocol-dependent,  and in general can be well captured by a linear form,
\beq
\varphi = -c \hat{T} + \varphi_{\rm J},
\label{eq:glass_EOS}
\eeq
where $c$ and $\varphi_{\rm J}$ depend on  $\hat{T}_{\rm g}$.
For the case $\hat{T}_{\rm g} = 0.033$, the parameters are $c= 1.59$ and $\varphi_{\rm J} = 0.682$ (see Fig.~\ref{fig:EOS}). 
Equation~(\ref{eq:glass_EOS}) can be used to estimate $\varphi$ from a given $ \hat{T}$ for the glass states, and vice versa.
For example, it gives a Gardner transition density $\varphi_{\rm G} = 0.671$ that corresponds to $T_{\rm G} = 0.0072$ obtained by the machine learning method (Fig.~4).

\section{Cumulants of caging order parameter}

\begin{figure}[ht]
\centerline{\includegraphics[width=0.5\textwidth]{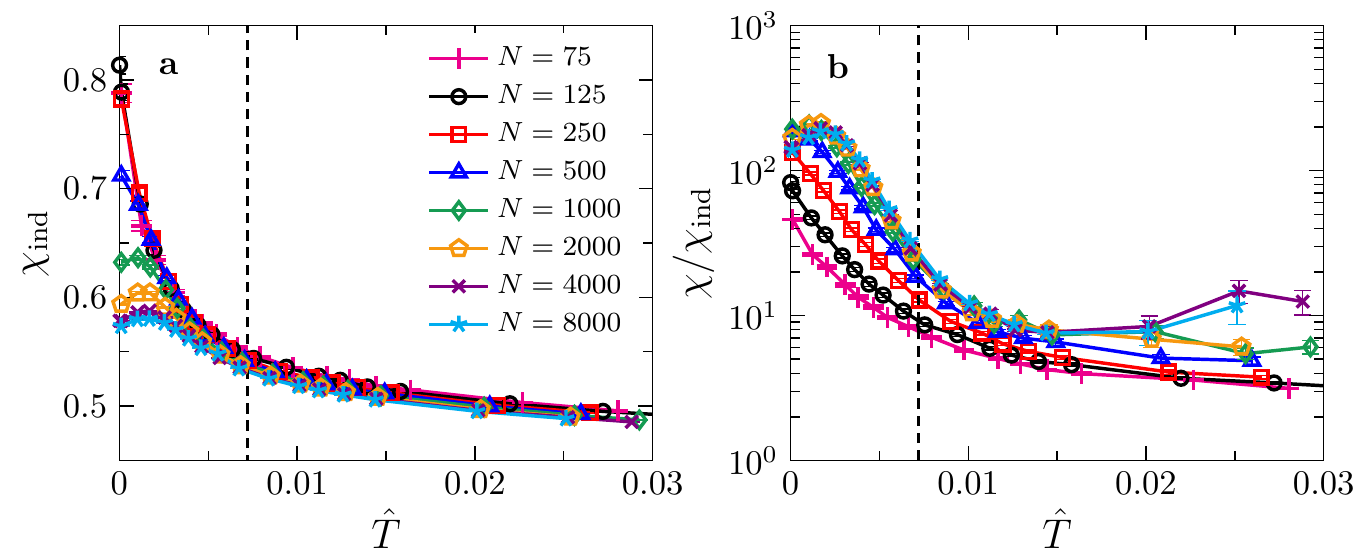} }
\caption{ (a) Individual caging susceptibility $\chi_{\rm ind}$ and (b) the ratio  $\chi/\chi_{\rm ind}$ as functions of $\hat{T}$, for $\Gamma = 10^{-4}$ and a few different $N$ ($N_{\rm r}=5$ and $N_{\rm s} = 1200$). The large fluctuations around $\hat{T}_{\rm g} = 0.033$ are caused by remaining $\alpha$-relaxations, which are suppressed at lower $\hat{T}$.
The Gardner transition temperature $\hat{T}_{\rm G} = 0.0072$ is marked by vertical lines. 
}
\label{fig:chi_single}
\end{figure}

\subsection{Sample-to-sample fluctuations}
\label{sec:sample_fluctuations}
In general, one can consider the total fluctuations of caging order parameter $\Delta_{\rm AB}$ over both glass replicas and equilibrium samples, by
\beq
\chi_{\rm tot} = N \overline{\left \langle \left(\Delta_{\rm AB} -  \overline{ \langle \Delta_{\rm AB} \rangle } \right)^2  \right \rangle},
\eeq
where $\langle x\rangle$ represents the average over $N_{\rm r}(N_{\rm r}-1)/2$ pairs of glass replicas obtained from the same equilibrium sample, and   $\overline{x}$ represents the average over $N_{\rm s}$ different equilibrium samples. The total susceptibility $\chi_{\rm tot}$ can be divided into two parts, $\chi_{\rm tot}  = \chi_{\rm r} + \chi_{\rm s}$, where
\beq
\chi_{\rm r} = N \overline{\left \langle \left(\Delta_{\rm AB} - \langle \Delta_{\rm AB} \rangle \right)^2  \right \rangle},
\eeq
and 
\beq
\chi_{\rm s} = N \overline{ \left(\left \langle\Delta_{\rm AB} \right \rangle - \overline{ \langle \Delta_{\rm AB} \rangle }  \right)^2  }.
\eeq
The first susceptibility $\chi_{\rm r}$
characterizes the fluctuations in different realizations of replica pairs, which is equivalent to the thermal fluctuations in long-time simulations. 
The second susceptibility $\chi_{\rm s}$
characterizes the fluctuations in different equilibrium samples (i.e., disorder). Although both susceptibilities are expected to diverge at the Gardner transition point in the thermodynamical limit, in small systems 
the sample-to-sample fluctuations near the critical point have complicated finite-size effects~\cite{berthier2016growing, charbonneau2015numerical}, which have been also noticed earlier  in spin glasses~\cite{parisi2012numerical}. For this reason, in the current study we only consider $\chi_{\rm r}$ (which is essentially equivalent to $\chi$ analyzed in the main text apart from normalizaiton), in order to minimize the effects of sample-to-sample fluctuations. 
We point out that the caging skewness $S$ and the Binder parameter $B$ measured here also correspond only to the thermal part (see {\it Materials and Methods}), while the caging skewness  measured in Refs.~\cite{berthier2016growing, charbonneau2015numerical} contains both thermal and disorder parts.

\subsection{Average single-particle caging susceptibility}

The average single-particle  caging susceptibility, or the {\it individual caging susceptibility}, $\chi_{\rm ind}$, is defined as,
$
\chi_{\rm ind} = \frac{1}{N} \overline{\sum_{i} \chi_i}
$ (see Fig.~\ref{fig:chi_single}a).
It is easy to show that the global susceptibility $\chi$  contains two parts,
$
\chi = \chi_{\rm ind} + \chi_{\rm corr},
$
where 
$
\chi_{\rm corr} = \frac{1}{N} \overline{\sum_{i\neq j} \langle u_i u_j \rangle}
$
is the contribution from the spatial correlations between single-particle order parameters (we have used $\langle u_{i} \rangle = \langle u_{j} \rangle =0$).
Figure~\ref{fig:chi_single}b shows that $\chi/\chi_{\rm ind} \sim \mathcal{O}(1)$ at high temperatures, suggesting an uncorrelated field of local order parameters.
The correlation grows quickly below the Gardner transition temperature $\hat{T}_{\rm G}$ as $\chi$ becomes a few hundred times larger than $\chi_{\rm ind}$.

\subsection{Robustness of finite-size and aging scalings of caging susceptibility with respect to parameters  and the aging protocol}
\begin{figure*}[ht]
\centerline{\includegraphics[width=0.8\textwidth]{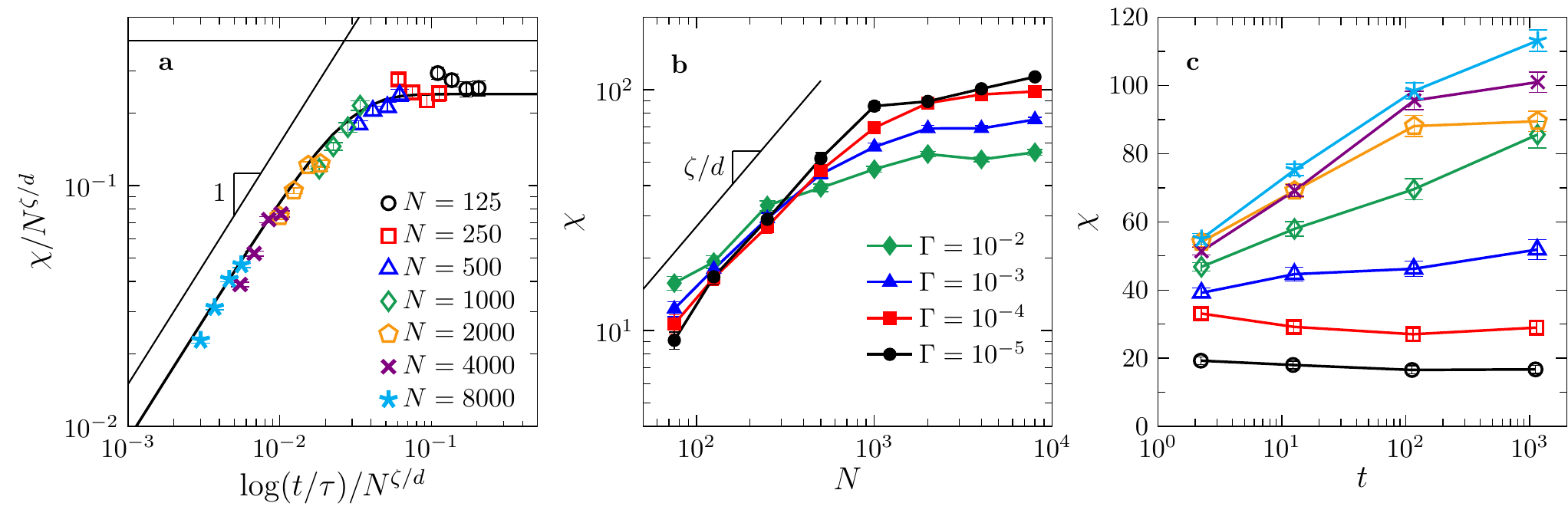} }
\caption{
Finite-size and aging scalings of caging susceptibility for $\hat{T}=0.00385$,  $N_{\rm r} = 10$ and $N_{\rm s} = 240$.
(a) Data collapsing according to Eq.~(6), for $125 \leq N \leq 8000$ and $\Gamma = 10^{-2}, 10^{-3}, 10^{-4},10^{-5}$, where $\zeta = 2.6$ and $\tau = 0.0016$ are used (same as in Fig.~1a).
The lines represent $\mathcal{F}(x\to \infty) \sim 1$, $\mathcal{F}(x\to 0) \sim x$, and an empirical fitting using the hyperbolic tangent function to guide the eye. 
(b) Susceptibility as a function of system size $N$, for a few different quench rates $\Gamma$. The line indicates $\chi \sim N^{\zeta/d}$. (c) Susceptibility as a function of quench time $t$, for a few different $N$.
The legend in (a) applies to both (a) and (c).
}
\label{fig:scaling_Nr10_Ns240}
\end{figure*}

According to the definition of $\chi$ (see {\it Materials and Methods} and Sec.~\ref{sec:sample_fluctuations}), the parameter $N_{\rm s}$  should only determine the statistical noise of the data, because $\chi$ only corresponds to  thermal fluctuations. 
On the other hand,  the value of $\chi$ is found to be dependent  on $N_{\rm r}$ (Fig.~1 and Fig.~\ref{fig:scaling_Nr10_Ns240}).
Nevertheless, Fig.~\ref{fig:scaling_Nr10_Ns240} shows that the scalings, Eqs.~(3-6), are robust with respect to $N_{\rm r}$, apart from the prefactors. 

\begin{figure}[ht]
\centerline{\includegraphics[width=0.5\textwidth]{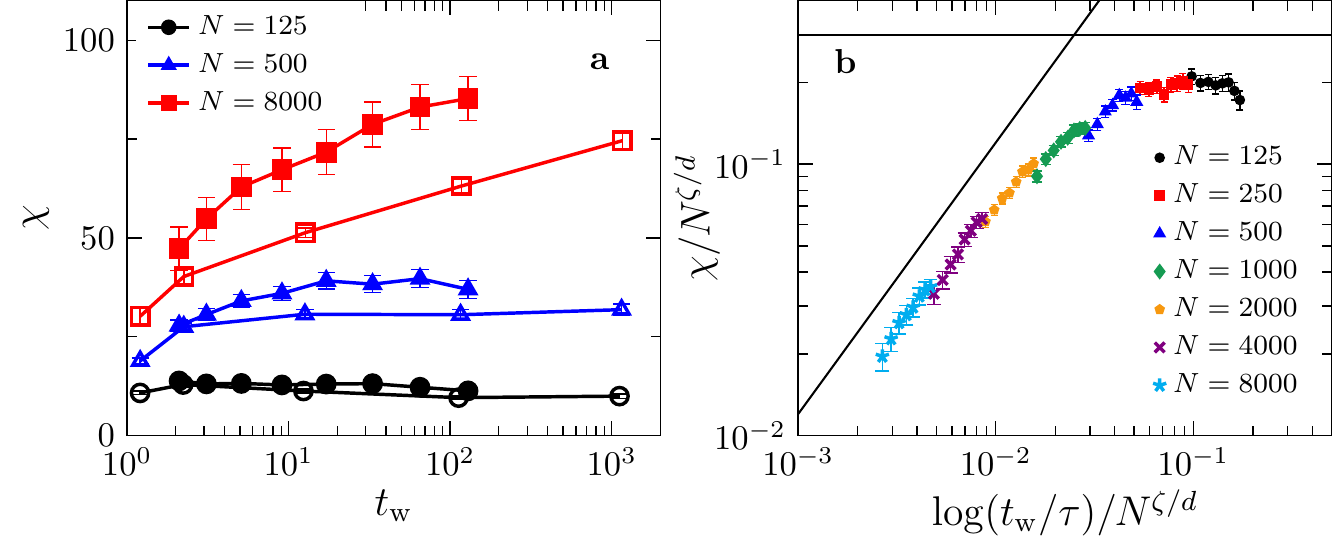} }
\caption{Caging susceptibility measured during isothermal aging after a rapid quench with $\Gamma = 0.01$.  (a) Susceptibility $\chi$ as a function of $t_{\rm w}$ for three different $N$ (filled symbols).
For comparison, the corresponding data in Fig.~1c, which are obtained using different $\Gamma$, are also plotted (open symbols). (b) Collapse of the data according to Eq.~(6), where the same parameters $\zeta = 2.6$ and $\tau = 0.0016$ as in Fig.~1a are used.} 
\label{fig:constP_aging}
\end{figure}

In the main text, aging is discussed as an effect for varying quench rate $\Gamma$ (or quench time $t$), where 
the system is compressed to a common reduced temperature $\hat{T}$ (or reduced  pressure $\hat{P} = 1/\hat{T}$).
The dependence of physical quantities (such as the susceptibility $\chi$) on the quench time $t$ (which is inversely proportional to $\Gamma$) is examined. 
Here we study another aging protocol -- isothermal aging, in order to  test the robustness of scaling Eq.~(6). In this  protocol, we first compress the system from $\hat{T}_{\rm g}$ to a target $\hat{T}$ with  a large  rate $\Gamma = 0.01$, {and set the waiting time $t_{\rm w} = 0$. We}
then relax the  system at a constant $\hat{T}$ and measure how the susceptibility evolves with the waiting time $t_{\rm w}$.
Thus this procedure mimics  isothermal aging (or equivalently isobaric aging  since our systems are hard spheres) after a rapid quench. 
Although the two aging protocols  give slightly different values of $\chi$, especially in large systems, the logarithmic growth behavior Eq.~(5) is robust  (Fig.~\ref{fig:constP_aging}a).
The data of $\chi$ obtained by both protocols can be  collapsed according to Eq.~(6), using the same parameters (Fig.~1a and Fig.~\ref{fig:constP_aging}b). Thus the scaling form and the exponent $\zeta$ are robust with respect to different aging protocols. The difference only presents in the pre-factors.

\subsection{Robustness of the critical scaling of caging skewness with respect to $N_{\rm r}$ }
\begin{figure}[ht]
\centerline{\includegraphics[width=0.5\textwidth]{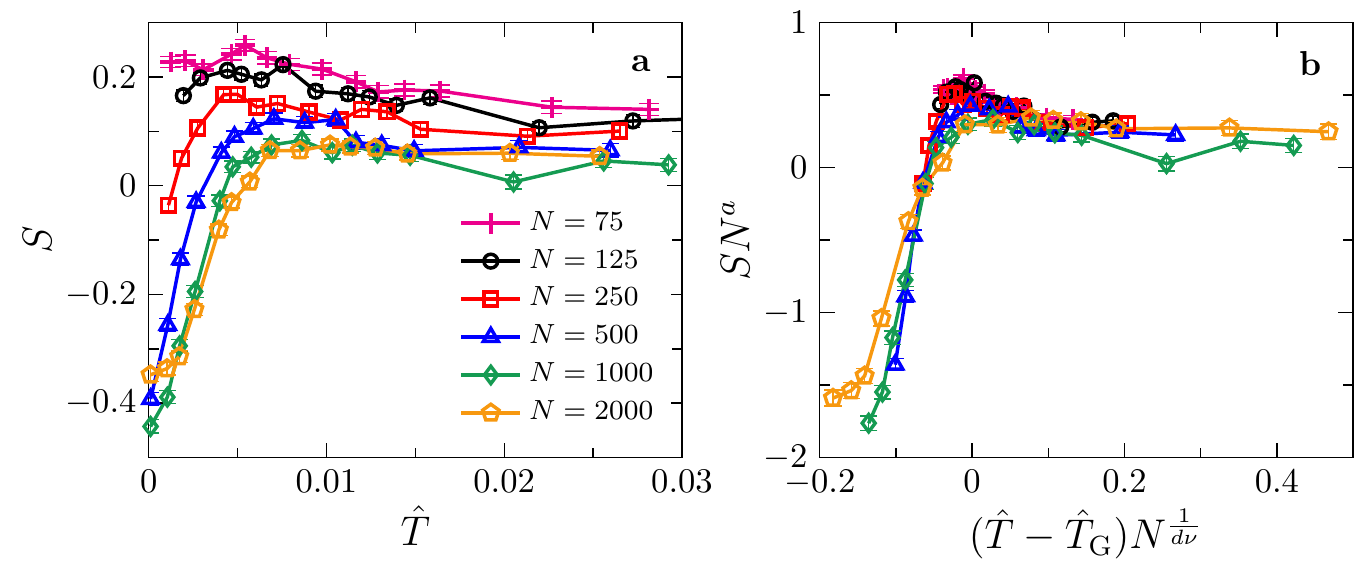} }
\caption{(a) Skewness $S$ as a function of $\hat{T}$ for a few different $N$, where $N_{\rm r} =5$ and $N_{\rm s} = 2400$ are used.  (b) Data collapsing according to the scaling ansatz $S N^a \sim \mathcal{S}\left[ (\hat{T} - \hat{T}_{\rm G})  N^{\frac{1}{d \nu}}\right]$, where $a=0.2$, $\hat{T}_{\rm G} = 0.0072$ and $\nu = 0.78$ as in Fig. 3d. 
}
\label{fig:skewness}
\end{figure}

Here we examine the influence of $N_{\rm r}$ on the  caging skewness. While the actual value of skewness slightly varies from $N_{\rm r} = 10$ (Fig.~3c) to $N_{\rm r} = 5$ (Fig.~\ref{fig:skewness}a), Fig.~\ref{fig:skewness}b shows that the proposed critical scaling $S N^a \sim \mathcal{S}\left[ (\hat{T} - \hat{T}_{\rm G})  N^{\frac{1}{d \nu}}\right]$ is more robust (except for the small deviations found for $N=2000 \approx N^*$).

\subsection{Binder parameter}
\begin{figure}[ht]
\centerline{\includegraphics[width=0.3\textwidth]{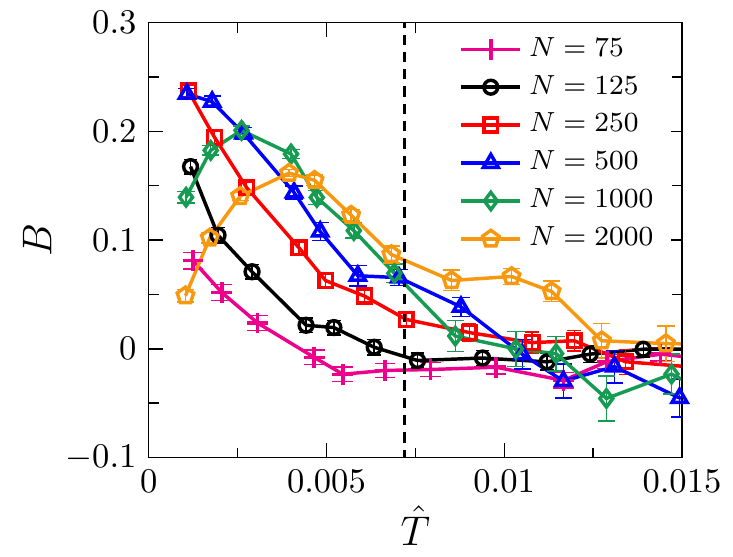} }
\caption{Binder parameter $B(\hat{T}, L)$ as a function of $\hat{T}$ for a few different $N$, obtained  from simulations using $N_{\rm r} = 10$, $N_{\rm s} = 2400$ and $\Gamma = 10^{-4}$.} 
\label{fig:binder}
\end{figure}

It is well known that, in the critical region of a standard second-order phase transition, the Binder parameter, which is the kurtosis of the order parameter distribution, satisfies a finite-size scaling $B(T, L) = \mathcal{B}[(T-T_{\rm c})L^{1/\nu}$], where $T_{\rm c}$ is the critical temperature. It means that the curves of $B(T, L)$ for  different $L$ should cross over at $T_{\rm c}$, which is commonly used to either examine the presence of a continuous phase transition, or to locate the critical point. However, it is difficulty to determine the phase transition using the Binder parameter for spin glasses in a magnetic field, due to strong finite-size corrections and the asymmetry of the order parameter distribution~\cite{ciria1993ahneida}. For the same reasons,
we do not observe a clear crossover in our data of  $B(\hat{T}, L)$ for the Gardner transition (see Fig.~\ref{fig:binder}). Note that the asymmetry of the order parameter distribution is clearly revealed  by the non-zero values of the skewness $S(\hat{T}, L)$ in Fig.~3c.

\section{Machine learning method}
\label{sec:ML}

\subsection{Designing input data}
\begin{figure}[ht]
\centerline{\includegraphics[width=0.35\textwidth]{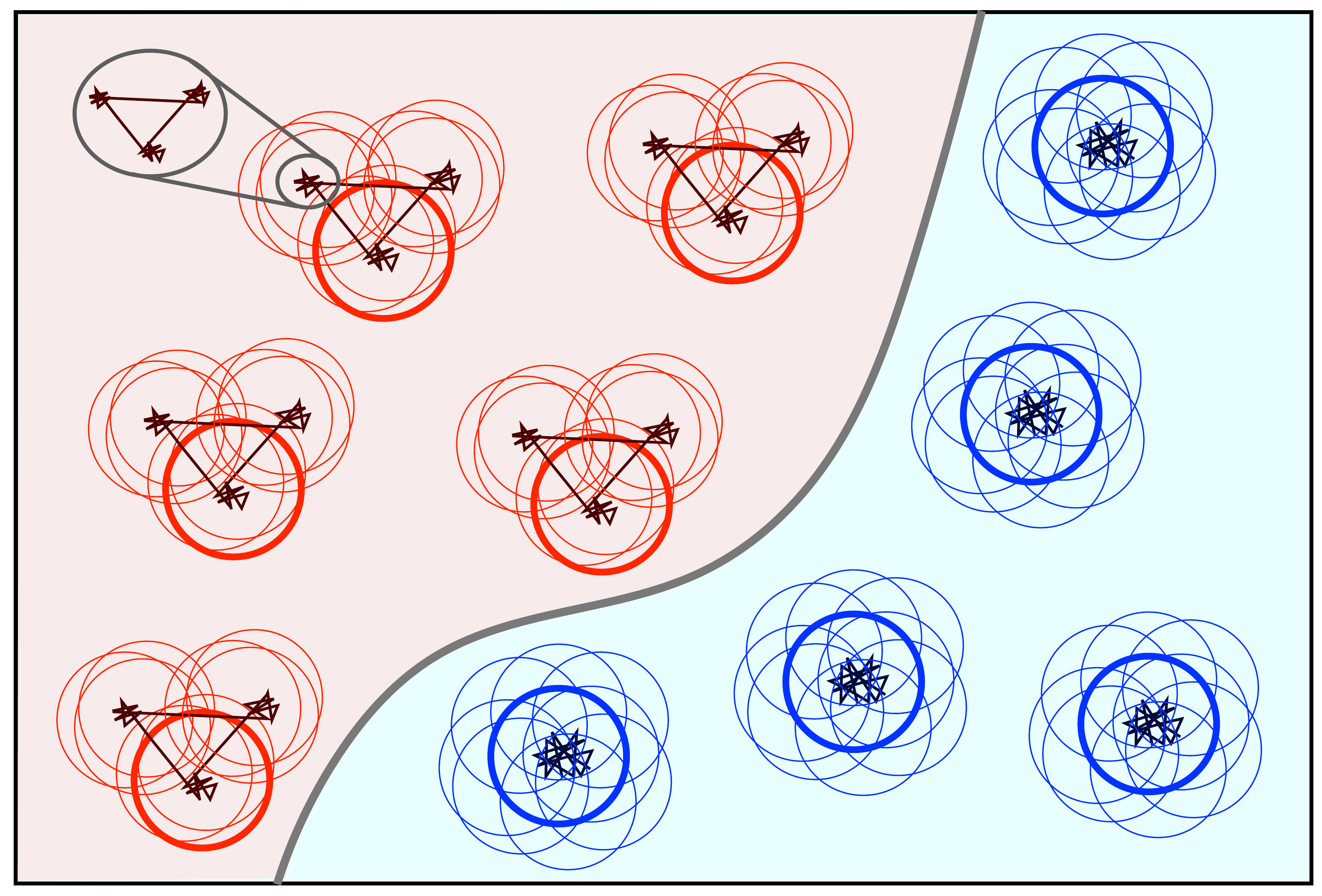} }
\caption{Illustration of particle vibrations in the Gardner phase. The vibrational features are demonstrated by the particle trajectories, and the organization of replicas (thin cycles). In the replica construction, the replicas of the same particle form a molecule~\cite{parisi2010mean}. The blue particles have simple vibrational cages with low susceptibility $\chi_i$, while the red particles have hierarchically split sub-cages with high $\chi_i$ (for simplicity, we only demonstrate two levels of split). The two kinds of particles are  organized  heterogeneously in space (red and blue areas).} 
\label{fig:chii_illustration}
\end{figure}

Particles in simple glass and Gardner phases have very different vibrational properties~\cite{charbonneau2017glass, berthier2016growing}. 
As illustrated in Fig.~\ref{fig:chii_illustration}, there are two kinds of particles in the Gardner phase. The first kind of particles (blue particles in Fig.~\ref{fig:chii_illustration}) have simple vibrational cages, while the second kind (red particles in Fig.~\ref{fig:chii_illustration}) have split sub-cages that are organized hierarchically. 
 The two kinds are clustered in space resulting in large vibrational heterogeneity~\cite{berthier2016growing}. 
 In contrast, only the first kind of particles exist in simple glasses. 
 

The above vibrational features were firstly revealed by the replica theory~\cite{charbonneau2014fractal}. In the theoretical construction, the original system $\{\vec{r}_1, \vec{r}_2, \ldots \vec{r}_N\}$ of $N$ particles are replicated $n$ times to form a molecular system~\cite{parisi2010mean}, $\{\vec{R}_1, \vec{R}_2, \ldots \vec{R}_N\}$, where each molecule consists of $n$ atoms, $\vec{R}_i = (\vec{r}_i^{1}, \vec{r}_i^2, \ldots \vec{r}_i^n)$. This ``replica trick" is realized in simulations by making $N_{\rm r}$ glass replicas from independent compressions of the same equilibrium sample (see {\it Materials and Methods}). In principle, one can use the full structure information of the molecular system  $\{\vec{R}_1, \vec{R}_2, \ldots \vec{R}_N\}$ as the input data for machine learning, and ask the algorithm to identify hidden features for different phases. However, this treatment would require a sophisticated design of the neural network (NN) architecture. In this study, based on the raw data  we construct a vector $\{\chi_1, \chi_2, \ldots \chi_N\}$ (see {\it Materials and Methods}).
As shown in Fig.~4a, the distribution $p(\chi_i)$ displays a single peak in the simple glass phase, suggesting that only one kind of particles exist. Moreover, the field of $\chi_i$ is distributed   homogeneously in space as expected (see Fig.~4c). In the Gardner phase, on the other hand, the distribution $p(\chi_i)$  exhibits two peaks. The particles in the left peak have simple vibrational cages, while those in the right peak have split vibrational cages with higher $\chi_i$. The particles belonging to different peaks are distributed heterogeneously in space as shown by the 3D plot in Fig.~4d. Therefore, the constructed vector $\{\chi_1, \chi_2, \ldots \chi_N\}$ well captures key particle vibrational properties, and with this treatment simple NN architectures are sufficient. Here we use a fully connected feedforward neural network (FNN) that has been shown to work for the phase identification in the Ising model~\cite{Carrasquilla2017Machine}.

We emphasize that it is the vibrational (or dynamical) features that can be used to distinguish between simple glass and Gardner phases. Structural ordering is not expected at the Gardner transition.
For this reason, it is impossible to learn the Gardner transition from static configurations $\{\vec{r}_1, \vec{r}_2, \ldots \vec{r}_N\}$. In principle, one can also try to construct the replicated molecular system from dynamical data, $\vec{R}_i = (\vec{r}_i(t_1), \vec{r}_i(t_2), \ldots \vec{r}_i(t_n))$, where $\vec{r}_i(t_k)$ is the position of particle $i$ at time $t_k$. 
This would require sufficiently long simulations in the Gardner phase such that particles perform enough hops to provide good sampling of sub-cages. 
However,  because hopping  in the Gardner phase is extremely slow (Fig.~1c), such long-time dynamical simulations are beyond present computational power.

It shall be also noted that, in the current design of input data, $\{\chi_1, \chi_2, \ldots \chi_N\}$, the information about spatial correlations between local caging order parameters is completely lost, since the particle coordinates  $\{\vec{r}_1, \vec{r}_2, \ldots \vec{r}_N\}$ are not included. 
The features of two phases are not learned  from 
the differences on caging heterogeneity  (see Fig.~4c-d). 
This point will be further discussed in Sec.~\ref{sec:shuffling}.

\subsection{Training and test data sets}
A total number of $N_{\rm s} \sim 2400$ equilibrium samples at $\hat T_{\rm g}$ are genearated by the swap algorithm.  
At each $\hat T < \hat T_{\rm g}$, $N_{\rm s}$ samples  of input data $\{\chi_1, \chi_2, ..., \chi_N\}$ are produced from quench simulations. 
The $N_{\rm s}$ samples  are divided into two sets. 
The training (or learning) set, which contains $N_{\rm s}^{\rm train}$ samples, is for training the FNN to learn the features of the simple glass and Gardner phases, outside  the blanking window $[\hat T_1, \hat T_2]$. The production set is for determining the phase transition, which is located inside $[\hat T_1, \hat T_2]$, blanked out during the training. Most previous applications of machine learning to identify phase transitions called the latter set the ``test'' set, following the machine learning terminologies. In digit recognition of machine learning, for example, the idea was to test the ability of a trained NN to identify unseen test set, which have known properties. Although we are not testing the trained FNN on the production set for accuracy, we still use the terminology of ``test'' set, to be consistent with the established protocols. The test set contains $N_{\rm s}^{\rm test}$ samples that are not included in the training set.

\subsection{Blanking window}
\begin{figure}[ht]
\centerline{\includegraphics[width=0.5\textwidth]{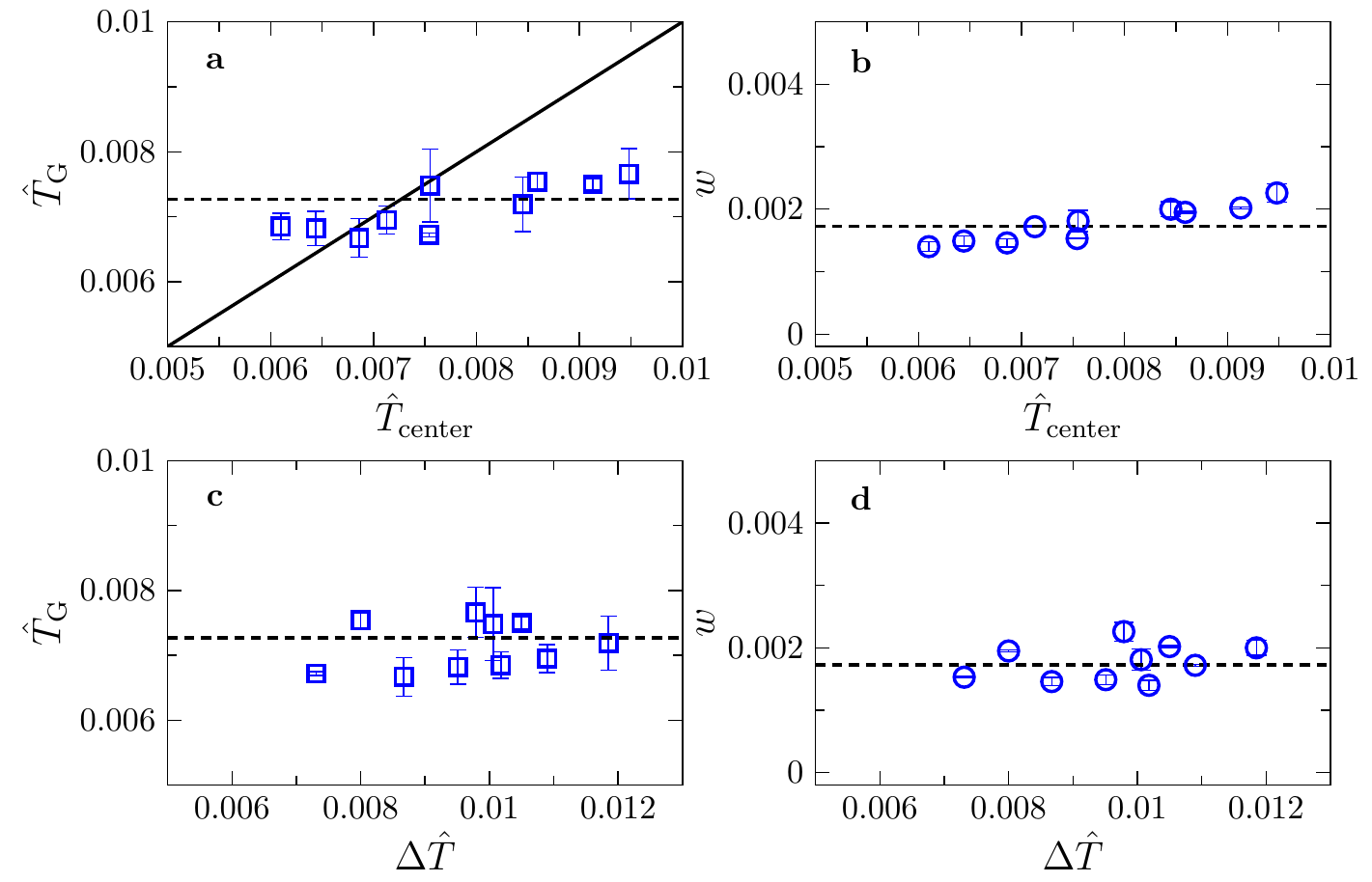} }
\caption{Examining the dependence  of machine learning predictions on the blanking window $[\hat{T}_2, \hat{T}_1]$. The FNN is trained using a few different combinations of $T_1$ and $T_2$, for data with $N=8000$ and $\Gamma = 10^{-4}$. The predicted  $\hat{T}_{\rm G}$ and $w$ are plotted as functions of $\hat{T}_{\rm center}$ and $\Delta \hat{T}$. The horizontal dashed lines mark the values  $\hat{T}_{\rm G}(N=8000) = 0.0073$ and $w(N=8000) = 0.0017$  used in Fig.~4 (obtained from $T_1 = 0.011$ and $T_2 = 0.0045$). 
The correlation between $\hat{T}_{\rm G}$ and $\hat{T}_{\rm center}$ is rather weak in (a), compared to the case in Fig.~\ref{fig:false_positive} for a false positive test, where the correlation is strong and close to $\hat{T}_{\rm G} = \hat{T}_{\rm center}$ (solid line).
} 
\label{fig:window}
\end{figure}

During supervised training, the glass states at $\hat{T} > \hat{T}_1$ are labeled as in the simple glass phase, while those 
at $\hat{T} < \hat{T}_2$ are labeled as in the Gardner phase (see {\it Materials and Methods}). 
The states within the blanking window $[\hat{T}_2, \hat{T}_1]$ are not used in the training.
Here we explain how to choose the parameters, $\hat{T}_{\rm center} = (\hat{T}_1 + \hat{T}_2)/2$  and   $\Delta \hat{T} = \hat{T}_1 -  \hat{T}_2$, for the blanking window.
Obviously, we should require $\hat{T}_{\rm G}$ to be inside of the blanking window, i.e., 
$\hat{T}_2 < \hat{T}_{\rm G}  < \hat{T}_{1}$.
Within this constraint, Fig.~\ref{fig:window}a-b show that  $\hat{T}_{\rm G}$ and  $w$ predicted by FNN (the two quantities plotted in Fig.~4) are weakly correlated to $\hat{T}_{\rm center}$.
To minimize the dependence on $\hat{T}_{\rm center}$, we choose $\hat{T}_{\rm center}$ to be in the range  [0.0062, 0.008], estimated from the minimal confusion principle that requires  the predicted $\hat{T}_{\rm G}$ to be as close as possible to the pre-assumed $\hat{T}_{\rm center}$ (ideally $\hat{T}_{\rm center} = \hat{T}_{\rm G}$). For such choices, both $\hat{T}_{\rm G}$ and  $w$  are independent of $\hat{T}_{\rm center}$ within the numerical error. Figure \ref{fig:window}c-d further show the independence of $\hat{T}_{\rm G}$ and  $w$ on the parameter  $\Delta \hat{T}$. Therefore, the choice of $\Delta \hat{T}$ is more flexible.


\subsection{Random shuffling}
\label{sec:shuffling}
Each input vector,
$\{\chi_1, \chi_2, ..., \chi_N\}$, has a particular ordering of the particle labels, an artifact kept from {\emph {off}}-lattice computer simulations of glasses, where a particle label needs to be created.
The shuffling of the elements in such a vector is identical to a simulated system with a different labeling order, which by itself is another valid sample. To remove the concept of labeling, here every original vector is duplicated $N_{\rm shuffle}$ times; each copy has a random ordering of the shuffled elements. Figure~\ref{fig:Ns}a shows how the machine learning results depend on $N_{\rm shuffle}$.

The shuffling is done here because the spatial correlations are already removed from the vector, $\{\chi_1, \chi_2, ..., \chi_N\}$, and become no further concerns. If one decides to use the raw data, $\{\vec{R}_1, \vec{R}_2, \ldots \vec{R}_N\}$, which contains particle correlations, care must be taken to use the machine learning approach; an {\emph {off}}-lattice simulation (e.g., liquids and glasses) produces no label-coordinate correlation and an {\emph {on}}-lattice simulation (e.g., Ising model and digitized hand-writing image) naturally maintains such a correlation. As discussed in Ref.~\cite{walters2019machine}, FNN is no longer the best choice to directly handle an 
{\emph {off}}-lattice dataset to explore spatial correlations. One should also check if random shuffling can be still applied since it can destroy the spatial correlations.

\subsection{Determining the number of training samples $N^{\rm train}_{\rm s}$}
\begin{figure*}[ht]
\centerline{\includegraphics[width=0.85\textwidth]{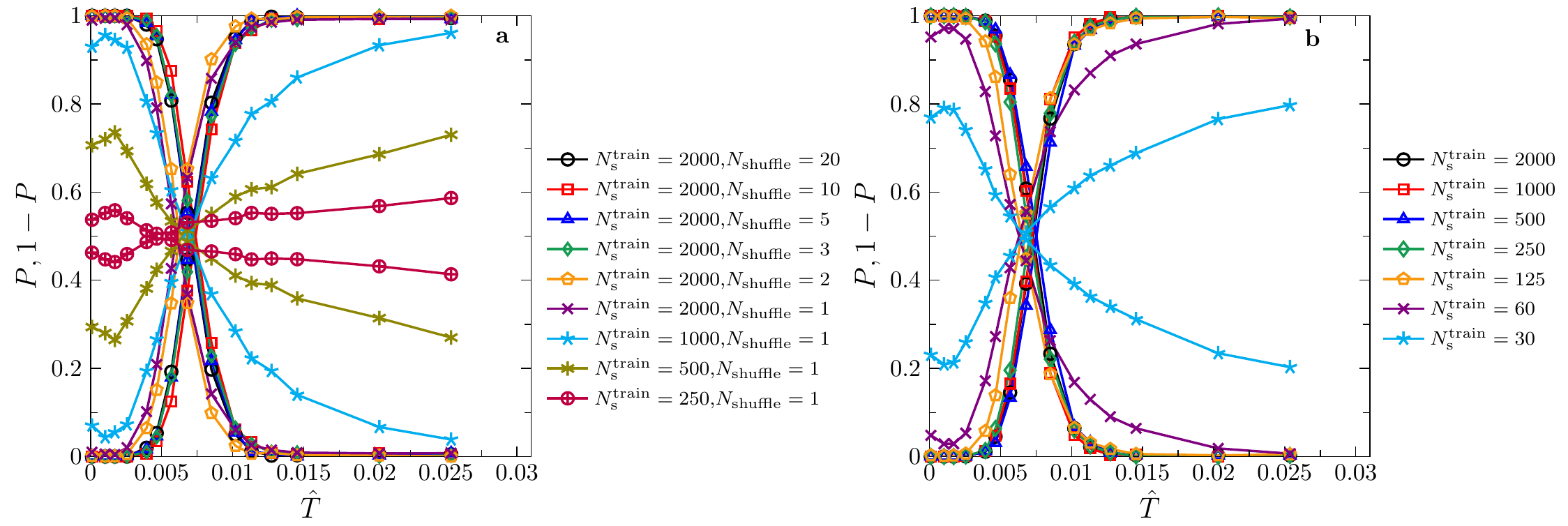} }
\caption{Influence of $N_{\rm shuffle} $ and $N^{\rm train}_{\rm s} $ on the probability $P$ obtained by the machine learning algorithm, with $N=2000$, $\Gamma = 10^{-4}$ and $N_{\rm r}=5$.
(a) The curves converge for $N_{\rm shuffle} \geq 3$ and $N^{\rm train}_{\rm s} =2000$. (b) The curves converge for  $N_{\rm shuffle} = 20$ and $N^{\rm train}_{\rm s} \geq 250$. } 
\label{fig:Ns}
\end{figure*}

It is well known that a machine learning method requires a large amount of training samples. To increase the size of training data set, we have introduced the trick of random shuffling. With this trick, generally the machine learning output converges when $N^{\rm train}_{\rm s} \gtrsim 250$ (for $N_{\rm shuffle} = 20$ random shuffles, see Fig. \ref{fig:Ns}b). The machine learning results presented in the main text are obtained using combinations of $N^{\rm train}_{\rm s}$ and $N_{\rm shuffle}$ such that $N^{\rm train}_{\rm s} \times N_{\rm shuffle} > 5000$.

\subsection{Independence of the number of clones $N_{\rm r}$}
\begin{figure}[ht]
\centerline{\includegraphics[width=0.3\textwidth]{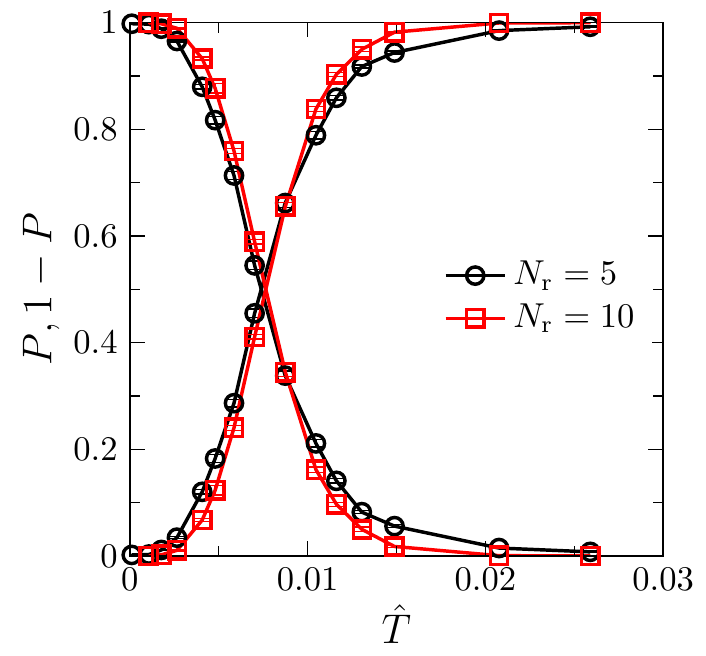} }
\caption{Influence of $N_{\rm r}$ on the probability $P$ obtained by the machine learning algorithm, with $N=500$, $\Gamma = 10^{-4}$, $N_{\rm shuffle}  = 20$ and  $N^{\rm train}_{\rm s} \sim 2000$.
} 
\label{fig:Nr}
\end{figure}

The input data of susceptibilities  $\{\chi_1, \chi_2, ..., \chi_N\}$ are calculated from $N_{\rm r}$ glass replicas (see {\it Materials and Methods}). Figure~\ref{fig:Nr} shows that the probability $P$ predicted by the machine learning algorithm is nearly independent of 
$N_{\rm r}$, when it is  increased from 5 to 10.


\subsection{A false positive test}

\begin{figure}[ht]
\centerline{\includegraphics[width=0.5\textwidth]{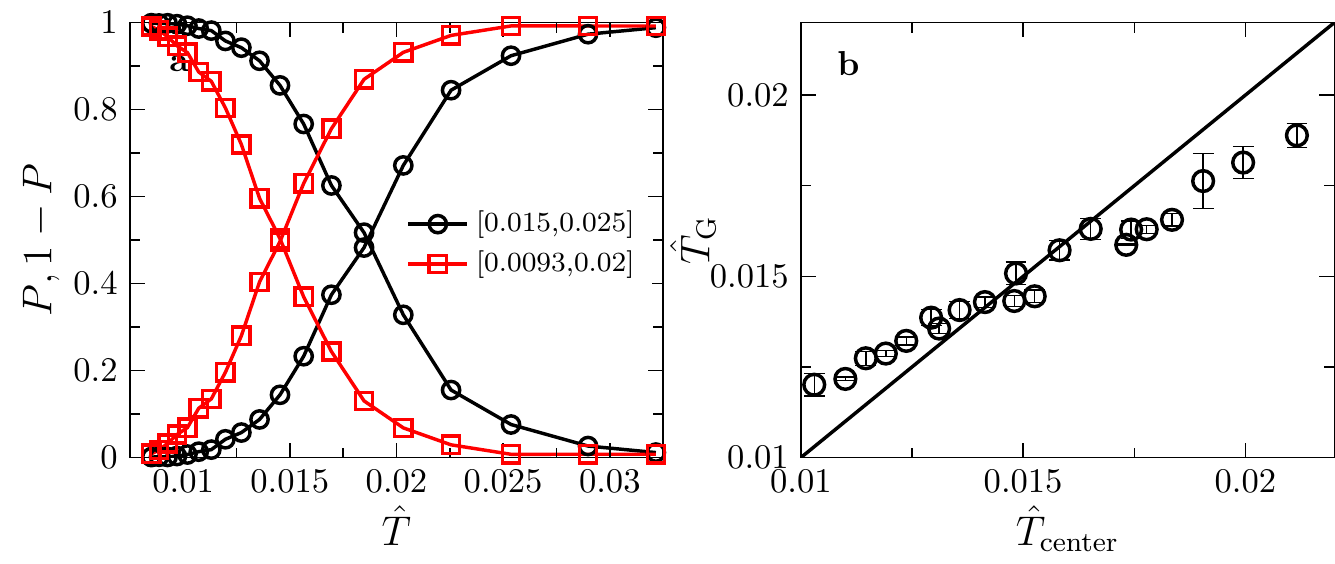} }
\caption{False positive test of the machine learning method for glass states in a temperature window $[0.0085, \hat{T}_{\rm g}]$. (a) Machine predicted $P$ and $1-P$ 
as functions of $\hat{T}$, for two different blanking windows. The curves are used to estimate a crossover point $\hat{T}_{\rm G}$ that is given by $P(\hat{T}_{\rm G}) = 0.5$.
(b) The predicted crossover point $\hat{T}_{\rm G}$ is strongly correlated to $\hat{T}_{\rm center}$. The line indicates  $\hat{T}_{\rm G} = \hat{T}_{\rm center}$.} 
\label{fig:false_positive}
\end{figure}

If all training and test samples belong to the same phase, would the machine learning method provide a false positive prediction  of a phase transition? 
To test this issue, we perform machine learning for glass states within a temperature window $[0.0085, \hat{T}_{\rm g}]$, which excludes the transition point $\hat{T}_{\rm G}=0.0072$.
Thus all training and test samples are in the simple glass phase. 
Clearly, if there is a phase transition and it can be correctly captured by the machine learning method, the predicted transition point should be independent of protocol parameters, as in Fig.~\ref{fig:window}.
On the other hand, Fig.~\ref{fig:false_positive} shows that the value of estimated $\hat{T}_{\rm G}$ is strongly correlated to  $\hat{T}_{\rm center}$, which is in sharp contrast to the case in  Fig.~\ref{fig:window}a, where $\hat{T}_{\rm G}$ is nearly independent of $\hat{T}_{\rm center}$. 
Therefore, one can unambiguously distinguish between the  case with a real phase transition (Fig. 4 and Fig.~\ref{fig:window}) and that 
simply corresponds a smooth  change within one phase (Fig.~\ref{fig:false_positive}).

\section{Data for quench rate $\Gamma = 10^{-2}$}
\begin{figure}[ht]
\centerline{\includegraphics[width=0.35\textwidth]{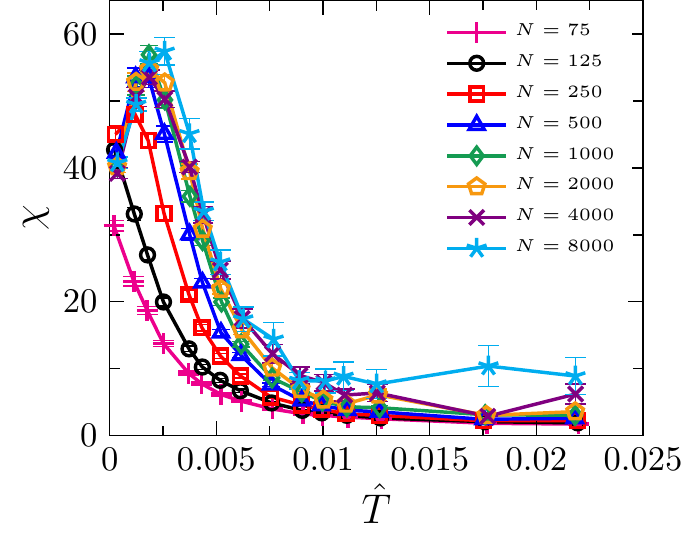} }
\caption{Caging susceptibility $\chi$ as a function of $\hat{T}$, for $\Gamma = 10^{-2}$ and a few different $N$. Data are obtained using $N_{\rm r} = 5$ glass replicas for each sample, and are averaged over  $N_{\rm s} = 480$ equilibrium samples.} 
\label{fig:Nstar}
\end{figure}

\begin{figure*}[ht]
\centerline{\includegraphics[width=0.9\textwidth]{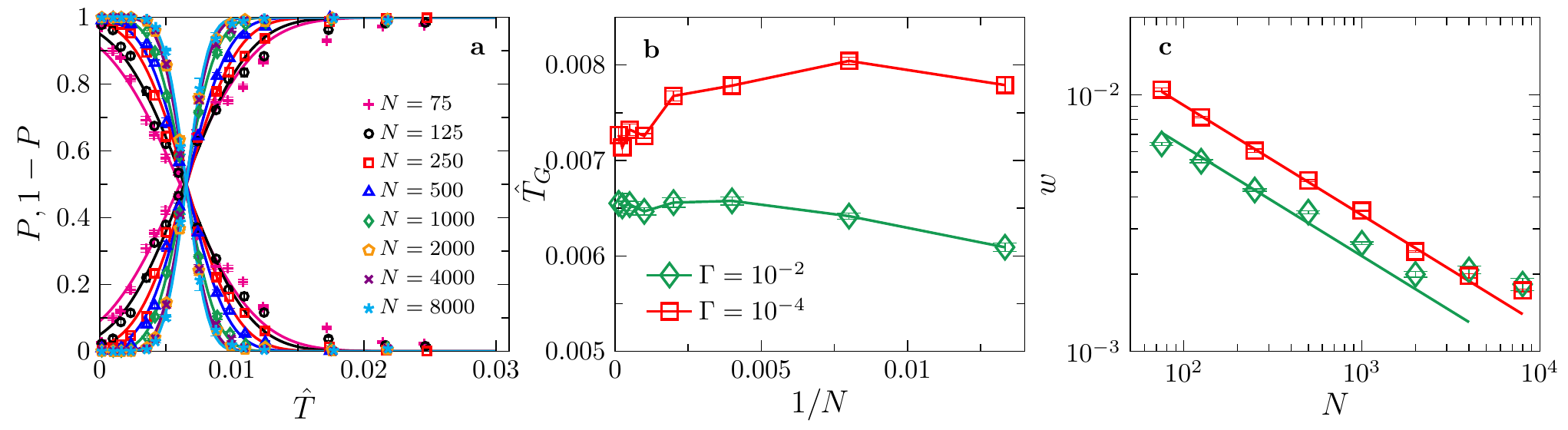}}
\caption{Machine learning results for $\Gamma = 10^{-2}$ ($N_{\rm r} = 5$, $N_{\rm s}^{\rm train}= 480$ and $N_{\rm shuffle}= 100$). (a) Probabilities $P(\hat{T}, N)$ and $1 - P(\hat{T}, N)$ as functions of $\hat{T}$. The lines  in (a) represent fitting to the form 
$P(\hat{T}, N) = \frac{1}{2} + \frac{1}{2}\erf \left\{\left[\hat{T} - \hat{T}_{\rm G}(N)\right]/w(N) \right\}$, where the fitting parameters $\hat{T}_{\rm G}(N)$ and $w(N)$ are plotted in (b) and (c), together with  corresponding results for $\Gamma = 10^{-4}$ from Fig.~4. The lines in (c) represent 
fitting according to the critical scaling $w(N) = w_0 N ^{-\frac{1}{d \nu}}$ within the range $N \leq N^*$ using $\nu = 0.78$, where $N^* \approx 1000$ for $\Gamma = 10^{-2}$ and $N^* \approx 2000$ for $\Gamma = 10^{-4}$. The rescaled plot $w(N)/w_0$ versus $N$ is presented in Fig.~4h  for both quench rates.} 
\label{fig:ML_Gamma1e-2}
\end{figure*}

To understand the influence of the quench rate $\Gamma$ on the criticality of the Gardner transition, additional simulations are performed by using a quench rate $\Gamma = 10^{-2}$. The data of susceptibility $\chi$ are plotted in Fig.~\ref{fig:Nstar}.
Comparing Fig.~\ref{fig:Nstar} to Fig.~3a where $\Gamma = 10^{-4}$, one can see that  $N^*$ shifts from $N^* \approx 1000$ for $\Gamma = 10^{-2}$ to $N^* \approx 2000$ for $\Gamma = 10^{-4}$. Here $N^*$ is the cutoff size above which the finite-size effect disappears. 
Accordingly, it is expected that the critical scaling regime $\omega(N) \sim N^{\frac{1}{d \nu}}$ around the transition point (Fig. 4), which only exists for $N \leq N^*$,  would shrink as $\Gamma$ increases. This is indeed confirmed by the machine learning results presented in Fig.~\ref{fig:ML_Gamma1e-2}c. The rescaled plot in Fig. 4h reveals the trend more clearly. The predicted $\hat{T}_{\rm G} = \hat{T}_{\rm G}(N \to \infty)$ is also slightly shifted with changing $\Gamma$ (Fig.~\ref{fig:ML_Gamma1e-2}b). Because $\hat{T}_{\rm G}$ increases with decreasing $\Gamma$, we do not expect $\hat{T}_{\rm G} \to 0$ in the zero quench rate limit.

\section{Comparing numerical critical exponents to theoretical predictions}
In Ref.~\cite{charbonneau2017nontrivial}, Charbonneau and Yaida predicted theoretically the critical exponents, $\nu$ and $\eta$, for the divergence of the correlation length and the power-law decay of the correlation function at the Gardner transition respectively, using a two-loop renormalization group (RG) calculation and the  Borel resummation 
based on a  three-loop calculation. Using the scaling relation, $2-\eta = \zeta$, we can also obtain the theoretical values of the exponent $\zeta$. The theoretical values are summarized in Table~\ref{table:exponents}. 
While two-loop and Borel resummation results are close to each other, the Borel resummation is expected to give more accurate values. Only the Borel resummation results are cited in the main text.

Ref.~\cite{berthier2016growing} estimated  $\eta \approx -0.32$ from fitting the power-law decay of the line-to-line correlation function obtained from simulation data at $\varphi = 0.67 \approx \varphi_{\rm G}$ for $\varphi_{\rm g} = 0.63$.
In this work, based on the machine learning approach,  we determine numerically $\nu = 0.78(2)$ (Fig.~4h). We also find power-law finite-size scaling regimes of the susceptibility data in the entire Gardner phase $\hat{T} \leq \hat{T}_{\rm G}$, and obtain values of the associated exponent, $\zeta = 1.5-3.0$, which weakly depends on the temperature $\hat{T}$ (Fig.~2b). In Table~\ref{table:exponents}, we compare these numerical measurements to theoretical predictions. 

\begin{table}[h]
\centering
\caption{Theoretical~\cite{charbonneau2017nontrivial} and numerical  critical exponents
for the Gardner transition in three dimensions. The numerical values of $\zeta$ are for $\hat{T} \leq \hat{T}_{\rm G}$, with $\zeta \approx 1.5$ at $\hat{T}_{\rm G}$.}
\begin{tabular}{ c |c| c | c}
\hline
\hline
& $\nu$ & $\eta$ &    $\zeta$ \\
\hline
two-loop theory~\cite{charbonneau2017nontrivial}   & 0.76 & -0.24     & 2.2  \\ 
 Borel resummation theory~\cite{charbonneau2017nontrivial}     & 0.85 & -0.13   & 2.1 \\
 simulation & 0.78(2)         & -0.32~\cite{berthier2016growing}            &1.5-3.0 \\
\hline
\hline  
\end{tabular}
\label{table:exponents}
\end{table}

\red{\section{Testing the machine learning method in a three-dimensional spin glass model}
The machine learning techniques used here have been well-documented in recent studies to identify the phase transitions in homogeneous critical systems, for example, the 2D Ising model~\cite{Carrasquilla2017Machine}.  The applicability of the techniques to study disordered systems, such as the case presented in this paper and the 3D spin-glass model, was previously unestablished. In this section, we show that the method can be used to pin-down the critical point and the correlation length critical exponent of the computer simulation data for the Edwards-Anderson spin glass model, which yields results that are consistent with known values. We also examine the finite-size and finite-time effects on the determination of critical parameters, similar to the case of the Gardner transition.}

\red{We consider an Ising Edwards-Anderson spin glass model, defined on a cubic lattice of linear size $L$ with periodic boundary conditions,  in $d=3$ dimensions. The Hamiltonian is 
\beq
H = - \sum_{\langle ij \rangle} J_{ij} s_i s_j, 
\eeq
where $s_i = \pm 1$ and $J_{ij}$ is a random variable that takes $\pm 1$ with equal probability. The summation is restricted to pairs $\langle i j \rangle$ of   nearest neighbors. Each instance of $\{J_{ij}\}$ is called a {\it sample}.
This model has been extensively studied, with a well established spin glass transition at $T_{\rm c} = 1.1019(29)$~\cite{baity2013critical, RuizLorenzo2020}. 
The value of the correlation length exponent 
is $\nu =  2.562(42)$~\cite{RuizLorenzo2020, baity2013critical}. }

\red{The model is simulated using Glauber Monte Carlo dynamics~\cite{zhou2015spin}. 
During one Monte Carlo step, which is defined as the unit of time, $L^3$ trials of (randomly chosen) spin flips   are attempted. 
Initial equilibrium configurations at $T+\Delta T$, where $\Delta T = 3.9$, are quenched to a target temperature $T$, with a fixed quench rate $\Gamma = dT/dt$. 
In order to examine the finite-time effects, we also apply an infinitely rapid quenching ($\Gamma = \infty$), and study how critical parameters depend on the waiting time $t_{\rm w}$ after the rapid quenching. 
If the waiting time $t_{\rm w}$ is shorter than the equilibrium time  $\tau$, the system is out-of-equilibrium.}

\begin{figure*}[ht]
\centerline{\includegraphics[width=0.9\textwidth]{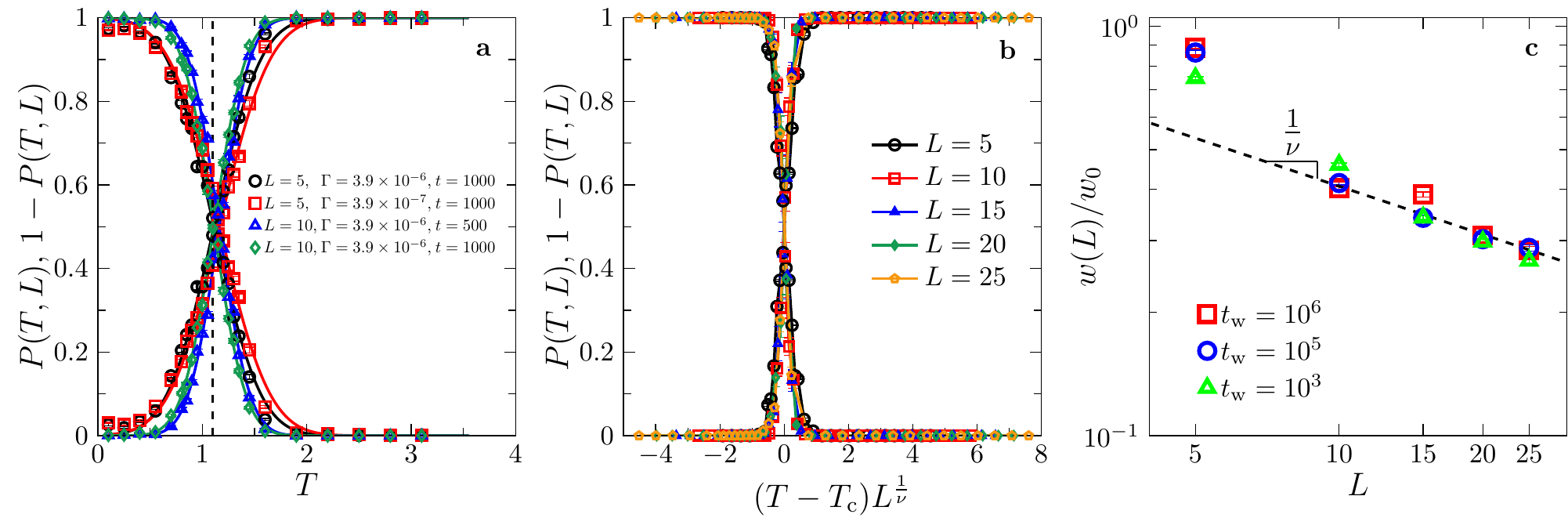} }
\caption{\red{Machine learning an Ising Edwards-Anderson spin glass model in three dimensions. (a) The probabilities  $P(T, L)$ 
of the paramagnetic phase and $1-P(T, L)$ of the spin glass phase,  obtained  from the machine learning method, are  
plotted as functions of temperature $T$,
for a few different $L$, $\Gamma$ and $\tilde{t}$ ($t_{\rm w} = 0$ in all cases). 
The dashed vertical line represents $T_{\rm c} = 1.1$~\cite{RuizLorenzo2020, baity2013critical}.
The solid lines represent fitting to an  empirical form
$P(T, L) = \frac{1}{2} + \frac{1}{2}\erf \left\{\left[T - T_{\rm c}(L)\right]/w(L) \right\}$, where $\erf(x)$ is the error function. 
The transition temperature (crossover point) $T_{\rm c}(L)$ and the width $w(L)$ are obtained from this fitting.
(b) Collapse of probability data  for the rescaled variable $(T- T_{\rm c}) L^{\frac{1}{\nu}}$, where $\nu = 2.56$~\cite{RuizLorenzo2020, baity2013critical} is used. The data are obtained from a rapid quenching ($\Gamma = \infty$) with $t_{\rm w} = 10^6$.
(c) Width $w(L)$ as a function of $L$, for $\Gamma = \infty$ and three different $t_{\rm w}$. The dashed line represents the scaling $w(L) = w_0 L^{-\frac{1}{\nu}}$, with $\nu = 2.56$.}} 
\label{fig:spin_glass}
\end{figure*}

\red{After quenching for time $\Delta T/\Gamma$ and waiting for additional time $t_{\rm w}$, we make $N_{\rm r} = 30$ {\it replicas}. These replicas share the same configuration 
at time $\Delta T/\Gamma + t_{\rm w}$, but evolve independently later on. Additional simulations are performed for a short period of time $\tilde{t}$ (note that the total time  is $\Delta T/\Gamma + t_{\rm w} + \tilde{t}$),  to obtain the final replica configurations that are used to calculate the {\it single-spin susceptibility},
\beq
\chi_i = \langle q_{AB}^2 \rangle - \langle q_{AB} \rangle^2,
\eeq
where $q_{AB} = s_i^A s_i^B$ is the overlap of the same spin $i$ in different replicas $A$ and $B$. 
In order 
to characterize ``vibrations'' in the spin glass phase, the time $\tilde{t} =1000$ is chosen to be 
shorter than equilibrium time $\tau$. 
 The vector $\{ \chi_i \}$ is used as the input data for the machine learning method. Once the input data sets are prepared following the above procedure, the same machine learning algorithm, 
as described in detail in {\it Materials and Methods} and Sec.~\ref{sec:ML}, can be applied. We use $N_{\rm s}^{\rm train} = 800$ 
samples for training, $N_{\rm s}^{\rm test} = 200$ additional samples for testing (prediction), and a blanking window 
$[T_2 = 0.55, T_1=2.2]$. The results are presented in Fig.~\ref{fig:spin_glass} and discussed in detail below.}

\red{We first show in Fig.~\ref{fig:spin_glass}a that the machine learning algorithm gives the  correct $T_{\rm c}$. 
For small systems ($L=5$ and 10), where it is easy to reach equilibrium for the chosen quench rate $\Gamma = 3.9 \times 10^{-6}$ ($t_{\rm w}=0$),
the crossover point given by the machine learning results is consistent with the standard value $T_{\rm c}\simeq1.1$, within the numerical precision.  The crossover point is nearly unchanged for a slower quench rate $\Gamma = 3.9 \times 10^{-7}$, or a smaller $\tilde{t}=500$.}


\red{Next, we use a finite-size analysis to examine if the machine learning method can provide a consistent prediction  of $\nu$ with previous studies. Using the known value $\nu \simeq 2.56$, our data of $P(T, L)$ for different $L$ can be nicely collapsed for the rescaled parameter $(T - T_{\rm c}) L^{\frac{1}{\nu}}$ (see Fig.~\ref{fig:spin_glass}b). Figure~\ref{fig:spin_glass}c further shows that the data of width $w(L)$ is consistent with the scaling $w(L) \sim w_0 L^{-\frac{1}{\nu}}$, except for the smallest systems.
In order to obtain data for larger $L$, here we have relaxed the requirement of equilibrium.
We find that the scaling in Fig.~\ref{fig:spin_glass}c is insensitive to the waiting time $t_{\rm w}$ after a rapid quenching. In fact, it is possible to estimate critical exponents from a finite-size analysis of data obtained from non-equilibrium systems.
Such an idea  has been already proposed in~\cite{lulli2016out}, although machine learning methods were not employed there. 
We will leave the application of the machine learning method to strictly equilibrated ensembles for  future studies, for which more sophisticated simulation algorithms, such as the parallel tempering method~\cite{hukushima1996exchange}, can be useful.}

\clearpage

\bibliography{MLG}

\end{document}